\documentclass[a4paper,fleqn,usenatbib]{mnras}

\usepackage{newtxtext,newtxmath}

\usepackage[T1]{fontenc}
\usepackage{ae,aecompl}

\usepackage{graphicx}	
\usepackage{amsmath}	
\usepackage{amssymb}	

\newcommand{\lboo}{$\lambda\,$Bo\"{o}}
\newcommand{\matd}{\mathrm{D}}

\title[Photospheric contamination]{Stellar Photospheric Abundances as a Probe of Disks and Planets}

\author[]{
Adam S. Jermyn\thanks{E-mail: adamjermyn@gmail.com}$^1$ \& Mihkel Kama$^{1}$\\
$^{1}$Institute of Astronomy, University of Cambridge, Madingley Road, Cambridge, CB3 0HA, United Kingdom \\
}

\date{Accepted XXX. Received YYY; in original form ZZZ}

\pubyear{2018}

\begin{document}
\label{firstpage}
\pagerange{\pageref{firstpage}--\pageref{lastpage}}
\maketitle

\begin{abstract}
Protoplanetary disks, debris disks, and disrupted or evaporating planets can all feed accretion onto stars.
The photospheric abundances of such stars may then reveal the composition of the accreted material.
This is especially likely in B to mid-F type stars, which have radiative envelopes and hence less bulk--photosphere mixing.
We present a theoretical framework (\texttt{CAM}) considering diffusion, rotation, and other stellar mixing mechanisms, to describe how the accreted material interacts with the bulk of the star.
This allows the abundance pattern of the circumstellar material to be calculated from measured stellar abundances and parameters ($v_{\rm rot}$, $T_{\rm eff}$).
We discuss the \lboo\ phenomenon and the application of \texttt{CAM} on stars hosting protoplanetary disks (HD~100546, HD~163296), debris disks (HD~141569, HD~21997), and evaporating planets (HD~195689/KELT-9).
\end{abstract}


\section{Introduction}

Accretion onto a star can originate in a protoplanetary or debris disk, in a disrupted or evaporating planet or companion, in the interstellar medium, or from material shed earlier by the star itself. The elemental composition of circumstellar material is of great interest but is hard to investigate, in particular for solids, elements of low absolute abundance, or elements whose major carrier species do not have easily observable energy transitions.

We present the Contamination by Accretion Method (\texttt{CAM}) -- physical equations which allow the photospheric composition of an accreting early-type star to be translated to the composition of circumstellar material. This includes a quantitative comparison of the main mixing mechanisms, including rotational, shear, thermohaline, and convective mixing as well as molecular diffusion. We neglect gravitational settling because this is important primarily for white dwarf systems which we do not consider here~\citep{1986ApJS...61..197P, Koesteretal2009}. The key observable is the stellar photospheric composition, (X/H)$_{\rm obs}$. This is a mixture of the stellar bulk composition, which we assume is a known reference value such as solar~\citep{Asplundetal2009}, and the accreted material. The accretion can originate in a circumstellar disk, an outflowing planetary atmosphere, or indeed an entire disrupted planet. These scenarios typically yield accretion rates from $10^{-14}$ to $10^{-7}\,$M$_{\odot}\,$yr$^{-1}$, and may have abundance patterns very different from, or identical to, the bulk of the star. Measuring the composition of the accreted mass fraction provides a new view on planet-forming and planetary material, complementary to observations explicitly targeting the circumstellar component.

\begin{figure}
	\includegraphics[width=0.47\textwidth, trim = 15 40 15 60]{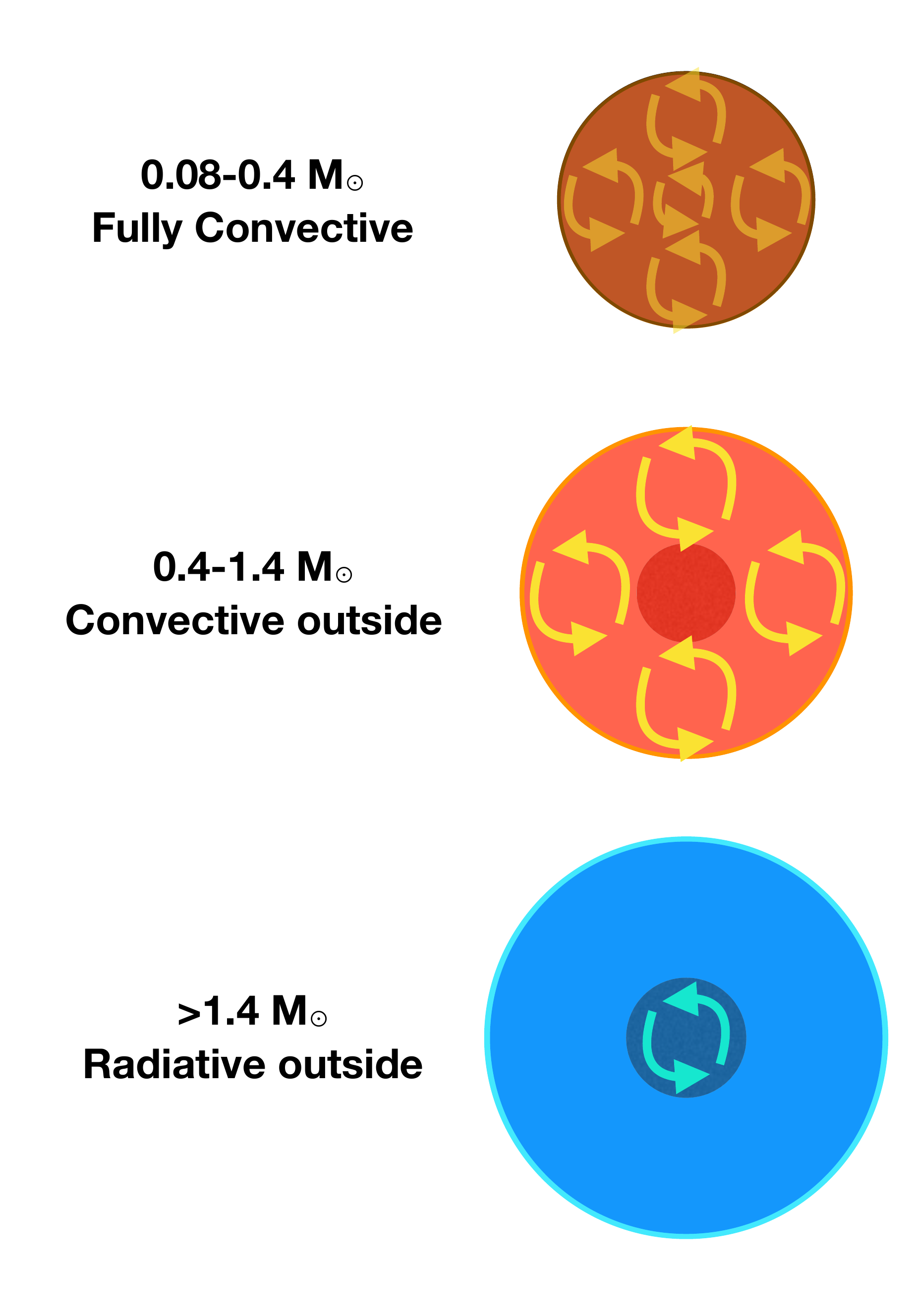}	
	\caption{The internal structure of low-mass ($M < 0.4 M_\odot$), intermediate-mass ($0.4 M_\odot < M <  1.4 M_\odot$) and high-mass ($M > 1.4 M_\odot$) stars are shown schematically with a focus on their mixing properties. Low-mass stars are fully convective, intermediate-mass ones like the Sun develop an outer convective layer, and high-mass ones lose this layer.}
	\label{fig:schematic}
\end{figure}

For a given stellar mass, rotation rate, and mass accretion rate, we show how to calculate the mass fraction, $f_{\rm ph}$, of freshly accreted, as opposed to stellar bulk, material in the photosphere. We then derive the composition of the recently accreted materials from this mass fraction. We discuss how to apply \texttt{CAM} to stars hosting protoplanetary and debris disks, and evaporating planets. The method works best on stars with masses of at least $1.4\,$M$_{\odot}$ because they have radiative envelopes, as shown in Fig.~\ref{fig:schematic}, such that accreted material is mixed down via slow non-convective processes~\citep{1991ApJ...372L..33C, CharbonneauMichaud1991, TurcotteCharbonneau1993, Turcotte2002}. For extremely high accretion rates, our method is also applicable to lower-mass stars where convection rapidly mixes accreted material with the bulk.

For the following, we define refractory elements as those having a condensation temperature $T_{\rm c}>300\,$K, constituting inter- and circumstellar dust, and volatiles as those having $T_{\rm c}<300\,$K, constituting ices and gases such as H$_{2}$O and CO. Furthermore, we recall that both for the Galactic interstellar medium as well as for a solar-composition mixture cooled to a few hundred kelvin, the mass ratio of gas to dust is $\Delta_{\rm g/d}=100$ \citep[e.g.][]{SnowWitt1996}.

The quintessential example of accretion-contaminated early-type stars are the $\lambda\,$Bo\"{o}tis objects. This subset of mid--F to B--type stars, roughly $2\%$~\citep{1998AJ....116.2530G}, have photospheres depleted in refractory elements by a factor of up to a few hundred. In practice, the \lboo\ occurrence rate is not known for spectral types earlier than B8 due to a lack of metal lines. Other types of chemically peculiar early-type stars, such as Ap and Am, can be produced by mechanisms intrinsic to the star \citep{Michaud1970} and do not show a correlation of abundances with condensation temperature. The \lboo\ anomaly is thought to require the preferential accretion of gas over dust -- i.e. volatile elements over refractories~\citep{VennLambert1990}. Many such stars, with a varying level of depletion, may be explained by one protoplanetary disk mechanism which excludes dust from the inner disk before it gets close enough to sublimate and accrete~\citep{Kamaetal2015}. Many if not all \lboo\ stars are, then, lighthouses of ongoing, recent, or extensive accretion of material with a gas-to-dust ratio different from the canonical $\Delta_{\rm g/d}=100$.

While \lboo\ stars are of interest to us, we henceforth consider more generally any early--type star accreting from a circumstellar material reservoir, including protoplanetary and debris disk hosts as well as stars with evaporating or disrupted planets. Given a stellar bulk metallicity (i.e. the elemental composition of the protostellar cloud from which the star originates), mass, age, and rotation rate, it is possible to constrain a combination of the mass accretion rate and composition of accreting material. The accretion rate can often be estimated from observations or predicted with models, which further constrains the composition. In this work, we present a theoretical framework for analysing accretion contamination and discuss specific cases which can be investigated with standard spectroscopic observations.

\section{Theory overview}

The picture we consider in this work is one of a star accreting material at a steady rate $\dot{M}$ from a circumstellar disc. If their compositions are different, the accreting material modifies the composition of the star as it accumulates. If the composition of the accreting material is constant during a given time period, then to describe the radial composition profile as a function of time it suffices to track the fraction $f$ of material at each point in the star which comes from accretion.

To make this task more straightforward we assume that the composition of the star is uniform along surfaces of constant column density $\Sigma$\footnote{These surfaces coincide with isobars in the case where the gravitational acceleration is uniform.}. This is equivalent to assuming that material which is accreted spreads out uniformly around the star in the angular direction on a timescale which is small compared to the vertical mixing time\footnote{This is not precisely true, but it should suffice for the level of precision available in both theory and observations. This is the case for two reasons. First, the mixing time increases rapidly with depth, so in most of the star it is the case that the vertical mixing time is long compared with the horizontal one. Secondly, the observable quantities of interest are, to leading order, linear in the amount of accreted material, and so if there is an angular distribution what is observed is an average over the star, which is precisely what this assumption provides.}. The reason to prefer column density over other measures is that it is a Lagrangian coordinate and so naturally incorporates the boundary conditions set by accretion.

\begin{figure}
	\includegraphics[width=1.0\columnwidth]{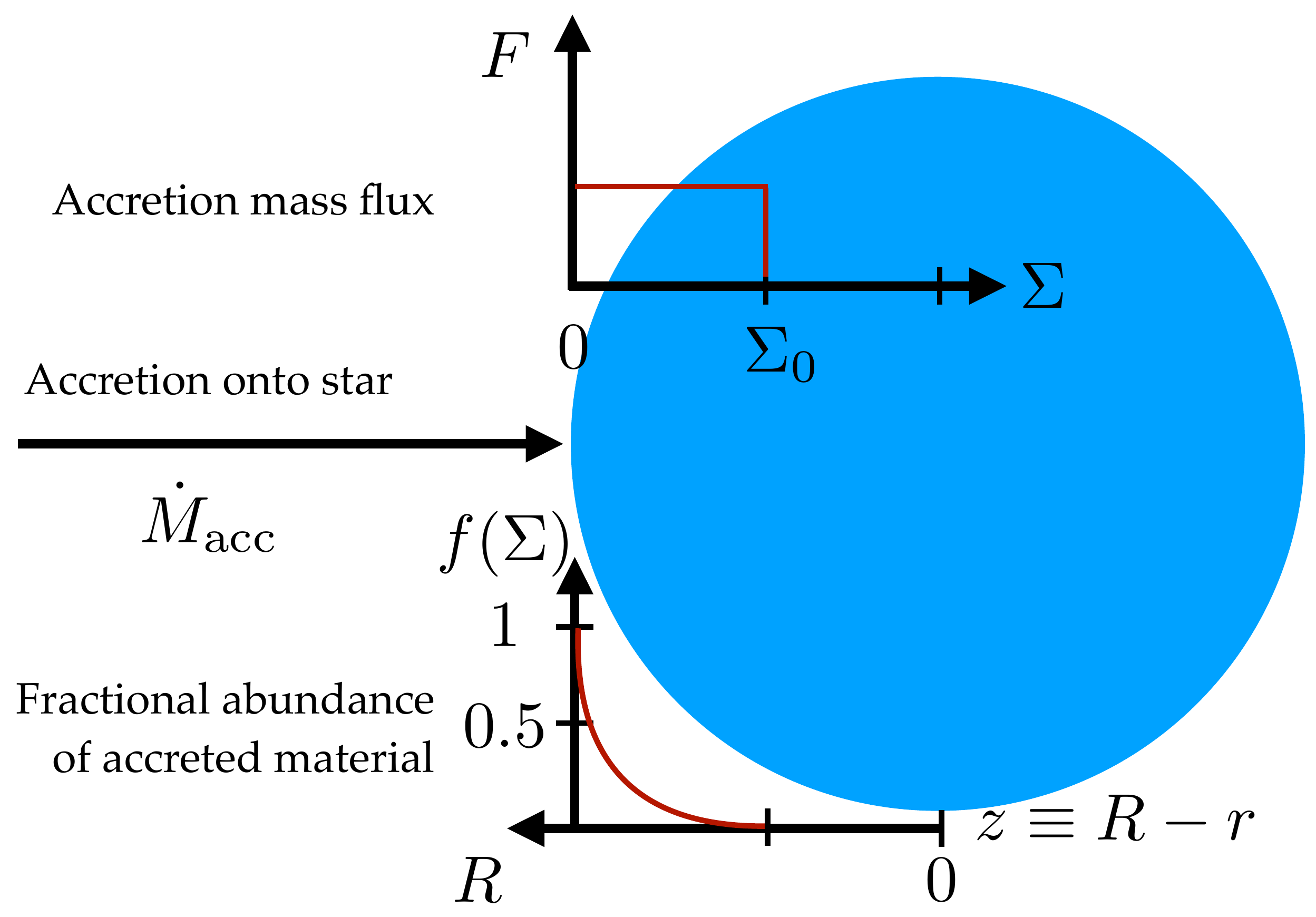}
	\caption{Illustration of the accretion contamination model. The radial coordinate $z$ runs inside-out, opposite to $r$ and the surface density $\Sigma$. The radial profile of $f$, the local fraction of accreted material, is given by Eq.~\ref{eq:fincrease}.}
	\label{fig:setup}
\end{figure}

The quantity we are after then is $f(\Sigma, t)$, where $t$ is the time since the accretion began.
As a further simplification we define $r(\Sigma)$ to be the mean radius associated with the corresponding surface of constant $\Sigma$. This allows us to work interchangeably with $r$ and $\Sigma$ without needing to impose spherical symmetry, which would disallow several mixing processes considered in Section~\ref{sec:mixingprocesses}. The model setup is visualized in Fig.~\ref{fig:setup}.

With these assumptions, the accreted material diffuses through the star according to
\begin{equation}
	\frac{\matd f}{\matd t} = -\frac{1}{\rho}\frac{\partial F}{\partial z},
	\label{eq:diff0}
\end{equation}
where $\rho$ is the density and
\begin{equation}
	z \equiv R - r,
\end{equation}
where $R$ is the surface radius of the star. Here
\begin{equation}
	F  = \rho D \frac{\partial f}{\partial z}
	\label{eq:flux0}
\end{equation}
is the flux of accreted material and $D$ is the spatially-varying diffusivity due to a variety of processes, including microscopic diffusivity, rotational mixing, convection and thermohaline mixing.
After some manipulations\footnote{See Appendix~\ref{appen:derivative}.}, equation~\eqref{eq:flux0} becomes
\begin{equation}
	\frac{\partial f}{\partial t} - \dot{\Sigma}\frac{F}{\rho^2 D} = -\frac{\partial F}{\partial \Sigma},
	\label{eq:diff3}
\end{equation}
where $\rho$ is the density and $\dot{\Sigma} \equiv \partial \Sigma/\partial t|_z$ is the column density accretion rate, given in terms of the mass accretion rate by
\begin{align}
	\dot{\Sigma} = \frac{\dot{M}}{4\pi R^2}.
\end{align}

As noted by~\citet{1991ApJ...372L..33C}, a full treatment of this system requires solving an unwieldy initial value problem.
In the case where the accreting material makes up a small fraction of the star the geometry is particularly straightforward; there is a single source on at the stellar surface and its interior is limitless for the purposes of the accretion -- even in heavily accreting protoplanetary disks, $M_{\star}/\dot{M}_{\rm acc}\sim 10^{7}\,$years.
It is possible then to approximate the mixing as the above authors and later~\citet{2002ApJ...573L.129T} did by a one-zone model, in which one tracks material entering and leaving a specific region of interest.
As we will find, however, the diffusivity and density change quite rapidly with depth and so this is not sufficient for our purposes.
Rather we will retain the spatial degree of freedom and make a somewhat less restrictive set of assumptions.

We begin by treating the system as being in instantaneous equilibrium so that
\begin{equation}
	\frac{\partial f}{\partial t} \ll \left|\dot{\Sigma}\frac{F}{\rho^2 D}\right|, \left|\frac{\partial F}{\partial \Sigma}\right|.
\end{equation}
This is motivated by the fact that in many cases the dominant effect of equation~\eqref{eq:diff3} is to transport material without changing $f$ locally.
This is obviously only an approximation but as we shall show in Appendix~\ref{appen:steady}, it holds quite well in the systems of interest.
With this approximation, equation~\eqref{eq:diff3} becomes
\begin{equation}
	\dot{\Sigma}\frac{F}{\rho^2 D} = \frac{\partial F}{\partial \Sigma},
	\label{eq:diff4}
\end{equation}
which is solved by
\begin{equation}
	F = A \dot{\Sigma} e^{\alpha \dot{\Sigma}},
	\label{eq:F0}
\end{equation}
where $A$ is a constant of integration,
\begin{equation}
	\alpha(\Sigma_{\mathrm{ref}},\Sigma) \equiv \int_{\Sigma_\mathrm{ref}}^\Sigma \frac{d\Sigma'}{\rho^2 D}
	\label{eq:alpha}
\end{equation}
and $\Sigma_{\mathrm{ref}}$ is an arbitrary reference depth.
Note that we assume that the thermal and pressure structure of the system is decoupled from the mixing fraction, such that this is time-independent.
Inserting this into equation~\eqref{eq:flux0} gives
\begin{equation}
	\frac{\partial f}{\partial \Sigma} = -\frac{A\dot{\Sigma} }{\rho^2 D} e^{\alpha \dot{\Sigma}} = -A\dot{\Sigma}  \frac{d\alpha}{d\Sigma} e^{\alpha \dot{\Sigma}}.
\end{equation}
Integrating yields
\begin{align}
	\int_0^{\Sigma} \frac{d\alpha}{d\Sigma} e^{\alpha \dot{\Sigma}} d\Sigma = \frac{1}{\dot{\Sigma}}\left[e^{\alpha(\Sigma_{\mathrm{ref}},\Sigma) \dot{\Sigma}}-e^{\alpha(\Sigma_{\mathrm{ref}},0) \dot{\Sigma}}\right]
\end{align}
and hence
\begin{equation}
	f(\Sigma) = f(0) - A\left[e^{\alpha(\Sigma) \dot{\Sigma}}-e^{\alpha(0) \dot{\Sigma}}\right],
\end{equation}
where we have dropped the reference column density argument for clarity.
The material at $\Sigma=0$ is what was just accreted, so $f(0) = 1$, hence
\begin{equation}
	f(\Sigma) = 1 - A\left[e^{\alpha(\Sigma) \dot{\Sigma}}-e^{\alpha(0) \dot{\Sigma}}\right].
	\label{eq:f0}
\end{equation}

Equation~\eqref{eq:f0} hides a complication, namely that there is nothing preventing $f$ from becoming negative at some $\Sigma$.
Indeed the integrand is monotonically increasing and $F_0$ is positive by definition, so this is a distinct possibility.
To prevent this the integration must halt at the depth $\Sigma_0$ at which $f(\Sigma_0)=0$, for here the instantaneous equilibrium approximation is clearly incorrect.
This depth parametrises the extent to which material has mixed into the star.
For $\Sigma > \Sigma_0$ we let $f(\Sigma) \approx 0$, as the solution must be monotonic in depth and yet cannot become negative.

The cutoff depth may be used to eliminate $A$. With this, equation~\eqref{eq:f0} becomes
\begin{equation}
	f(\Sigma) = 1 - \frac{e^{\alpha(\Sigma)} - e^{\alpha(0)}}{e^{\alpha(\Sigma_0)} - e^{\alpha(0)}}.
	\label{eq:f2}
\end{equation}
This defines a family of solutions, one for each possible value of $\Sigma_0$.
We can relate these solutions to the time evolution by means of mass conservation.
Let the total accreted column density be
\begin{equation}
\Sigma_\mathrm{acc} \approx \int_0^t \frac{\dot{M}_{\mathrm{acc}}}{4\pi r^2} dt,
\end{equation}
where this is approximate because we have neglected the variation in $r$ with depth.
This column density must equal the integrated column density of accreted material, so
\begin{equation}
\Sigma_\mathrm{acc} = \int_0^t \dot{\Sigma} dt \approx \int_0^{\Sigma_0} f(\Sigma) d\Sigma.
\label{eq:mass}
\end{equation}
This defines $\Sigma_{\mathrm{acc}}$ as a function of $\Sigma_0$.
Because this is a monotonic relation it may be inverted to define $\Sigma_0$ as a function of $\Sigma_{\mathrm{acc}}$ and to thereby define $f$ as a function of time.

To invert this relationship there are two cases we must consider: either $d\alpha/d\Sigma$ asymptotes to zero as $\Sigma \rightarrow \infty$ or it does not.
In the former case we find that
\begin{equation}
	t \approx \frac{h^2}{D(\Sigma_0)}
	\label{eq:increase}
\end{equation}
(see Appendix~\ref{appen:asymptote} and equation~\ref{eq:increasingAppen}),
where
\begin{equation}
	h \equiv -\frac{dr}{d\ln p}
	\label{eq:h}
\end{equation}
is the pressure scale height, which we evaluate here at $\Sigma_0$, and $p$ is the pressure.
This may be interpreted physically as the material diffusing to the column density at which the diffusion timescale matches the accretion time, which is a generic feature of diffusion problems with a localised source and dynamic diffusivity increasing away from the source.
The associated material fraction is well-approximated by
\begin{equation}
	f(\Sigma) \approx \min\left(1, \frac{h^2 \dot{\Sigma}}{D \Sigma}\right).
	\label{eq:fincrease}
\end{equation}

While the case where $d\alpha/d\Sigma$ remains non-zero does not arise physically with any of the transport processes we consider here~\footnote{See Section~\ref{sec:mixingprocesses}.}, it is still worth examining for completeness.
In this case we find (see equation~\eqref{eq:decreasingAppen}) that
\begin{equation}
	\Sigma_{\mathrm{acc}} \approx \Sigma_0,
	\label{eq:decrease}
\end{equation}
with
\begin{equation}
	f(\Sigma) \approx 1.
\end{equation}
This is because the dynamic diffusivity is decreasing inward, and so material just piles up at the surface.
At the interface between the two cases there is more complex behaviour (e.g. when $\alpha$ diverges slowly), but the qualitative picture is well-captured by equations~\eqref{eq:increase} and~\eqref{eq:decrease}
As shown in Appendix~\ref{appen:steady}, both cases satisfy our initial assumption that the time derivative may be neglected.

If the accretion suddenly halts the system continues to mix, but now $\dot{\Sigma} = 0$ and the boundary condition on $f(0)$ is replaced with a condition on the surface flux, namely that
\begin{equation}
	F(0) = 0,
\end{equation}
subject to the equation
\begin{equation}
	\frac{\partial f}{\partial t} = -\frac{\partial F}{\partial \Sigma}.
\end{equation}
Once more taking the instantaneous equilibrium approximation we find that $F$ is a constant, in this case zero, such that
\begin{equation}
	\frac{\partial f}{\partial \Sigma} = 0,
\end{equation}
and hence $f$ is a constant as well.
The boundary between the region in which the old solution applies and that in which the new one applies sits at the depth $\Sigma_\mathrm{b}$, such that
\begin{equation}
	f(\Sigma_\mathrm{b})_- = f(\Sigma_\mathrm{b})_+,
	\label{eq:sigb}
\end{equation}
where the subscripts $-$ and $+$ refer respectively to the region at greater depth than $\Sigma_\mathrm{b}$ and that at shallower depth.
Along with conservation of mass this yields a time evolution relation, namely that the amount of accreted material in the region which has adapted to the new surface condition is
\begin{equation}
	\Sigma_{\mathrm{acc}+} = \int_0^{\Sigma_\mathrm{b}} f(\Sigma) d\Sigma = \Sigma_{\mathrm{acc}} - \int_0^{t} F(\Sigma_{\mathrm{b}})_{-} dt,
	\label{eq:massConsA}
\end{equation}
where time is now measured since the accretion ended.
This gives
\begin{equation}
	f(\Sigma < \Sigma_{\mathrm{b}}) = 1 - \frac{1}{\Sigma_{\mathrm{acc}}}\int_0^{t} F(\Sigma_{\mathrm{b}})_{-} dt,
	\label{eq:fsb}
\end{equation}

\begin{figure}
\includegraphics[clip=,width=0.97\columnwidth]{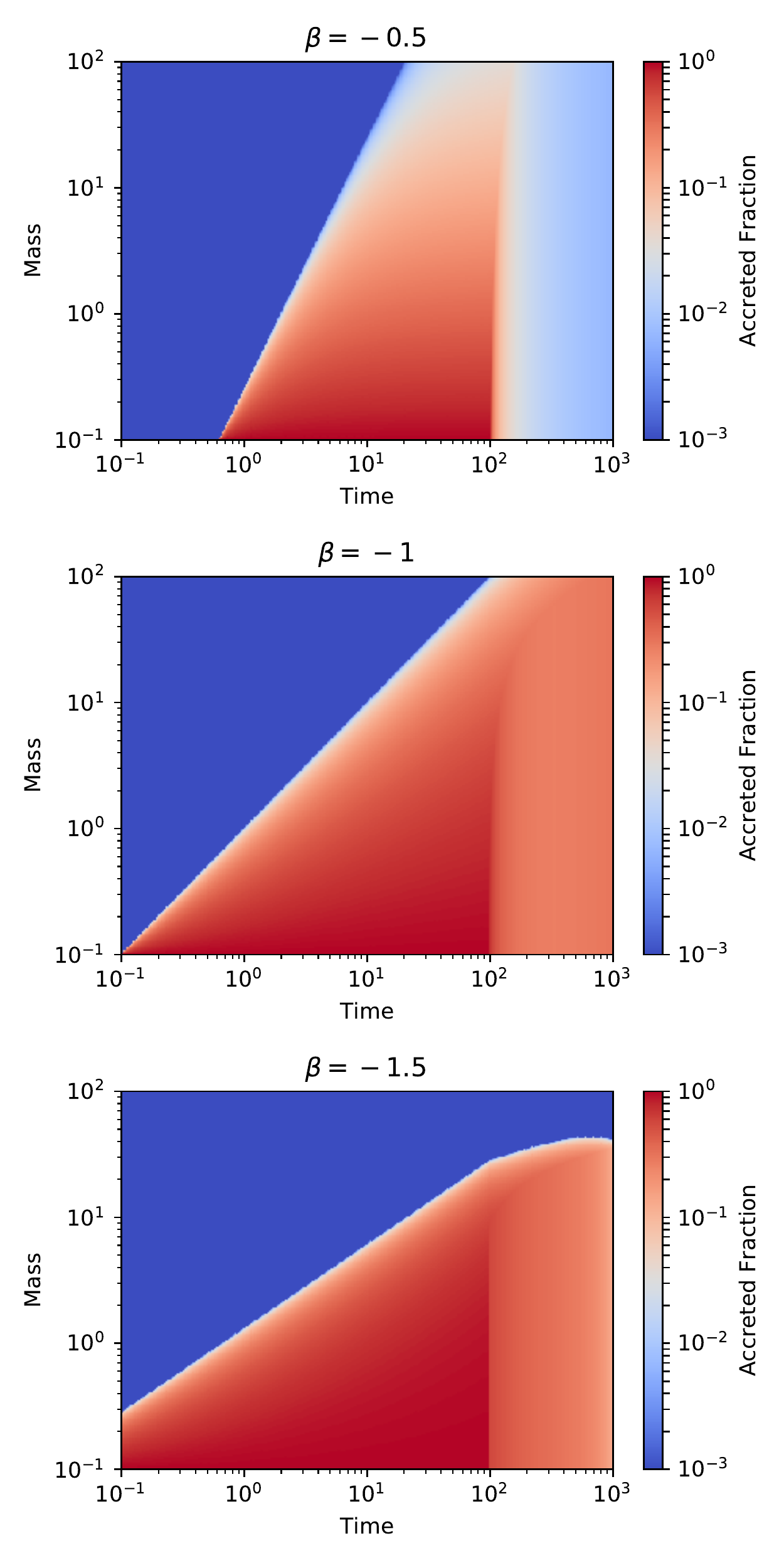}
\caption{The accreted fraction $f$ is shown as a function of mass coordinate, defined to be zero at the surface, and time for three stars with different power-law diffusivities ($\beta = -0.5$, $\beta=-1$, and $\beta=-1.5$ respectively from top to bottom). These models were computed using the instantaneous equilibrium assumption for both $\Sigma < \Sigma_{\mathrm{b}}$ and $\Sigma > \Sigma_{\mathrm{b}}$. In both cases the units are arbitrary, and were chosen such that $h=D_0=1$ and $\dot{\Sigma}=0.1$. Accretion begins at time $t=0$ and ends at $t=10^2$. In both cases the timescale for mixing away from the surface after accretion ends is set by the depth to which the material had reached, which in turn is set by the timescale over which material was accreted.}
\label{fig:kip}
\end{figure}

As we only ever find ourselves in the case given by equation~\eqref{eq:fincrease}, we combine this with equation~\eqref{eq:sigb} to find that at depths for which $\dot{\Sigma} h^2/D(\Sigma_\mathrm{b}) \Sigma_\mathrm{b} < 1$,
\begin{equation}
	\frac{\dot{\Sigma} h^2}{D(\Sigma_\mathrm{b}) \Sigma_\mathrm{b}} = 1 - \frac{1}{\Sigma_{\mathrm{acc}}}\int_0^{t} F(\Sigma_{\mathrm{b}})_{-} dt,
\end{equation}
where $\dot{\Sigma}$ is the accretion rate prior to it halting.
Assuming that $\rho \propto h$ and that $D$ obeys a power-law in $\Sigma$~\footnote{See Appendix~\ref{appen:asymptote} for a discussion of these approximations.} we find
\begin{equation}
	\Sigma_{\mathrm{b}} \approx \Sigma_{\mathrm{acc}} \frac{h^2}{D(\Sigma_{\mathrm{b}}) (t_{\mathrm{acc}} - t)},
\end{equation}
where $t_{\mathrm{acc}}$ is the time over which the system accreted.
Letting $\Sigma_{\mathrm{b}} = \Sigma_0$ and using equation~\eqref{eq:increase} we find
\begin{equation}
	\Sigma_{\mathrm{b}} \approx \Sigma_{\mathrm{acc}} \frac{t_{\mathrm{acc}}}{t_{\mathrm{acc}} - t}.
	\label{eq:sigbt}
\end{equation}
Thus we see that the timescale over which the photospheric composition returns to its pre-accretion state is essentially the timescale over which the accretion occured in the first place.
This behaviour may be seen in Fig.~\ref{fig:kip}, which shows the evolution of the accreted fraction with time for three stars which differ only in the behaviour of the diffusivity.
More specifically, we have taken
\begin{equation}
	D = D_0 \Sigma^{\beta},
\end{equation}
and varied $\beta$ between the three stars.
What is shown is the solution to equations~\eqref{eq:f2},~\eqref{eq:increase},~\eqref{eq:sigb},~\eqref{eq:massConsA}, and~\eqref{eq:fsb}.
In both cases accretion is turned on at $t=0$ and off at $t=10^2$.
In the first case, with $\beta = -0.5 > -1$, the star rapidly mixes material to incresing depths, and when the accretion is turned off the material disperses into the star in a timescale comparable to the accretion time.
In the second case, with $\beta=-1$, the marginal (e.g. logarithmic) scaling of $f$ with time (in the small-$f$ regime) that was found in Appendix~\ref{appen:asymptote} is apparent.
Finally in the second case, with $\beta = -1.5 < -1$, the diffusion front progresses much more slowly.
In each case when the accretion is turned off the surface rapidly relaxes to match the peak accreted fraction, which it then tracks as the material slowly disperses.
In particular, the phenomenology is precisely what we have described in equation~\eqref{eq:sigbt}, with the timescale for dispersal is being set by the depth to which the material had reached, which in turn is set by the timescale over which material was accreted.
The notable exception to this is if $D_0$ changes after the accretion halts, in which case the new timescale is
\begin{equation}
	t' = t \frac{D_0}{D_0'}.
	\label{eq:dnew}
\end{equation}

\section{Mixing Processes}\label{sec:mixingprocesses}

In the stars of interest there are five basic mixing processes which could play a role:
\begin{enumerate}
\item Molecular Diffusion
\item Rotational Mixing
\item Shear Mixing
\item Thermohaline mixing
\item Convection
\end{enumerate}

To first order these effects contribute additively to the diffusivity, and so each may be considered independent of the others.
In this section we examine each of these in turn.
We then compare their relative magnitudes as a function of the stellar parameters and discuss the extent to which they ought to contribute to mixing in accreting A-type stars.

\subsection{Molecular Diffusion}

Molecular diffusion, also known as microscopic diffusion, is the process whereby random thermal motions of particles in a fluid mix material within the fluid.
The fluids of interest are ionised hydrogen-dominated plasmas, so the diffusivity is
\begin{equation}
	D \approx 5.2\times 10^{-15} \left(\ln \Lambda\right)^{-1} \left(\frac{T}{\mathrm{K}}\right)^{5/2} \left(\frac{\rho}{\mathrm{g/cm^3}}\right)^{-1} \mathrm{cm^2 s^{-1}}
	\label{eq:visc0}
\end{equation}
\citep{1956pfig.book.....S}, where
\begin{equation}
\ln\Lambda =
\begin{cases} 
      -17.4+1.5\ln T-0.5\ln \rho & T<1.1\times 10^5 \mathrm{K} \\
      -12.7+\ln T-0.5\ln\rho & T>1.1 \times 10^5 \mathrm{K}
   \end{cases}
\end{equation}
and $T$ and $\rho$ are measured in C.G.S.K. units when they appear in a logarithm.
For simplicity we take $\ln \Lambda = -10$ and ignore both thermal and magnetic corrections to this factor.
Inserting this into equation\ \eqref{eq:visc0} and rescaling with a representative temperature and density we find
\begin{equation}
	D \approx 5\times 10^{-5} \left(\frac{T}{10^4\mathrm{K}}\right)^{5/2} \left(\frac{\rho}{0.1\mathrm{g/cm^3}}\right)^{-1} \mathrm{cm^2 s^{-1}}.
	\label{eq:visc1}
\end{equation}
When this mode of mixing dominates the diffusivity is a power-law with $\beta=-1$ (see Appendix~\ref{appen:asymptote}).

\subsection{Rotational Mixing}

Rotational mixing is a result of the Eddington-Sweet circulation.
While this manifests as a meridional flow, near the surface it has radial scale $z$ and velocity
\begin{equation}
	u \approx \left(\frac{F_\star}{p}\right) \frac{\Omega^2 R}{g}
\end{equation}
\citep{1929MNRAS..90...54E}, where $\Omega$ is the angular velocity, $F_\star$ is the heat flux and $p$ is the pressure.
The radial velocity is offset from this by a factor of $h/R$, so the effective radial diffusivity is
\begin{equation}
	D \approx u_r h \approx h\left(\frac{h}{R}\right)\left(\frac{F_\star}{p}\right)\frac{\Omega^2 R}{g} = \left(\frac{z}{R^2}\right)\left(\frac{F_\star}{p}\right)\frac{u_\mathrm{rot}^2}{g},
	\label{eq:rotdiff}
\end{equation}
where $u_\mathrm{rot}$ is the rotation speed at the surface, which is typically of order tens to hundreds of kilometres per second.
This prescription is quite similar to that found by~\citet{1992A&A...253..173C} in a more detailed analysis.

Equation\ \eqref{eq:rotdiff} gives rise to $\beta=-1$.
The abundances are thus enhanced over what we expect for $\beta > -1$ by a factor of $\ln (D_0 t / h^2)$\footnote{See equation~\eqref{eq:fBetaMinus1} and the preceding discussion.}, and so varies logarithmically in time rather than asymptoting to a fixed value.
For the purposes of our calculations below we neglect this logarithmic factor, but simply note that it could cause the real abundances to exceed those that we predict by a factor of several when this is the dominant mixing mechanism.

\subsection{Shear Mixing}

Shear mixing is due to differential rotation or other shears present in the star.
Non-rotational shears may come from meridional circulations, but in that case they cannot produce more mixing than the circulation itself because they have the same fundamental length-- and time--scales as the circulation~\citep{2009pfer.book.....M}.
Thus we need only consider shears due to differential rotation.

There are two main sources of differential rotation in these systems, namely the angular momentum brought to the surface of the star by the infalling material and the long-term buildup of angular momentum due to meridional circulations.
While in systems with very strong accretion the former may be severe, in most cases the accretion magnetically truncates outside the star~\citep{2014IAUS..302...91L}.
When this occurs the infalling material is brought into corotation and so does not create a surface shear.
Rather the angular momentum is transported via the magnetic field, which may spread it through a large volume of the star.
As such we neglect the shears due to accreting material.

Along similar lines, the long-term buildup of angular momentum is expected to be quite large in some cases~\citep{1992A&A...265..115Z, 2000A&A...361..101M, doi:10.1111/j.1365-2966.2011.18766.x}, but astereoseismic measurements appear to run counter to this prediction~\citep{2015MNRAS.454.1792K,0004-637X-788-1-93, refId01111, refId011}.
There have also been suggestions that more exotic angular momentum transport mechanisms are at work near the surfaces of stars~\citep[see e.g.][]{1999ApJ...520..859K}.
Given the uncertainties involved in this physics we neglect this effect in our analysis, but note that it could change our results should there be strong shears in the stars of interest.

\subsection{Thermohaline Mixing}

When the accreting material is metal-poor mixing due to an unstable molecular weight gradient ought not to occur.
This is because such a situation leads to a stable molecular weight gradient, which is the opposite of what is needed for the molecular gradient mixing mechanism.
One might worry that a stable weight gradient would hinder other mixing mechanisms, but this effect should not be significant.
While the effect of accreting metal-poor material on the spectrum of the star may be profound, its impact on the molecular weight gradient is small, especially if the ratio of hydrogen to helium in the accreting material matches that in the star.
As such we are justified in neglecting all effects associated with molecular weight gradients for such scenarios.

By contrast when the accreting material is metal-rich relative to the star it causes an unstable molecular weight gradient.
This produces a diffusivity of the form
\begin{equation}
D \approx C K \left|\frac{d\ln\mu}{d\ln p}\right|
\end{equation}
\citep{1972ApJ...172..165U}, where $\mu$ is the molecular weight, $C \approx 10^3$ is a constant and
\begin{equation}
	K = \frac{4acT^3}{3\rho^2 \kappa c_\mathrm{p}}.
\end{equation}
It is worth noting that there is considerable uncertainty in the constant $C$.
Reported values inferred from observations include $658$~\citep{2007A&A...467L..15C}, $667$~\citep{2010A&A...522A..10C} and approximately unity~\citep{1980A&A....91..175K}.
Numerical as well as more theoretically-motivated studies show similar disagreement, with values including $2$~\citep{2010A&A...521A...9C}, of order $10$~\citep{2011ApJ...728L..29T}, $658$~\citep{1972ApJ...172..165U} and $1294$~\citep{2013ApJ...768...34B}\footnote{Note that there is some variation in the precise conventions used to define $C$ which may result in differences of factors of several between studies, including in some cases distinguishing between the microscopic thermal and material diffusivities~\citep{2013ApJ...768...34B}, but comparative studies which standardise around consistent definitions indeed find considerable disagreement in the literature~\citep{2011ApJ...728L..29T}.}.
We use $C = 10^3$ for this study because it provides good agreement with observations, but this remains a significant source of uncertainty.

Using
\begin{equation}
	\frac{df}{dz} = \frac{F_0}{\rho D}
\end{equation}
we find that
\begin{align}
\frac{d\ln\mu}{d\ln p} &= \frac{d}{d\ln p}\ln\left[\mu_\star (1-f) + f \mu_\mathrm{acc}\right]\\
&= \frac{\mu_\mathrm{acc}-\mu_\star}{\mu_\star (1-f) + f \mu_\mathrm{acc}}\frac{df}{d\ln p}\\
&= \frac{\mu_\mathrm{acc}-\mu_\star}{\mu_\star (1-f) + f \mu_\mathrm{acc}}h\frac{df}{dz}\\
&= \frac{\mu_\mathrm{acc}-\mu_\star}{\mu_\star (1-f) + f \mu_\mathrm{acc}}\frac{h F_0}{\rho D}.
\end{align}
Thus
\begin{equation}
D \approx \sqrt{\frac{h F_0 C K}{\rho} \left|\frac{\mu_\mathrm{acc}-\mu_\star}{\mu_\star (1-f) + f \mu_\mathrm{acc}}\right|}.
\label{eq:materialD}
\end{equation}
This scales like $\rho^{-3/2}$, so $\beta=-3/2$ and hence the dynamic diffusivity actually decreases with depth.
As a result this mechanism is rapid in the photosphere and becomes slow deeper down.
As in the case of rotational mixing, because $\beta \leq -1$ there is a logarithmic enhancement in abundances as a function of time which we neglect for the purposes of our abundance calculations because it is generally small.

Because this mechanism depends on $f$ and $F_0$ in principle it should be incorporated by solving the differential equations defining $F$ and $\Sigma$ anew.
Instead of this we simply solve equation~\eqref{eq:fIncreasing} consistently with equation~\eqref{eq:materialD}.
This amounts to making a local calculation of the mixing near the photosphere.
Because the photospheric abundance is primarily set by mixing processes near the photosphere this ought to be a good approximation, even if its naive extrapolation into the interior is more problematic.

In the opposite case, in which a star accretes material which is much lighter than the bulk composition, the material gradient acts to suppress certain kinds of turbulent mixing, most notably convection.
However surface convection zones are somewhat superadiabatic, so we are justified in generally ignoring this effect.

\subsection{Convective Mixing}

Convection transports heat by means of bulk turbulent mixing of material.
The associated diffusivity is typically approximated via Mixing Length Theory as
\begin{align}
	D \approx h v_\mathrm{c}
	\label{eq:mlt}
\end{align}
\citep{1958ZA.....46..108B}, where $h$ is as defined in equation\ \eqref{eq:h} and $v_\mathrm{c}$ is the convection speed.
When convection is efficient (i.e. nearly adiabatic) the latter may be approximated as
\begin{align}
	v_\mathrm{c} \approx \left(\frac{F_\star}{\rho}\right)^{1/3}
	\label{eq:vc}
\end{align}
\citep{1992isa..book.....B}, where $F_\star$ is the heat flux.
At the photosphere proper this is not a good approximation, as a significant fraction of the flux there is carried by radiation, but below this point the approximation typically becomes very good\footnote{The error in this approximation falls exponentially in optical depth at low depths before hitting a (typically low) floor set by the radiative diffusivity.}.
Furthermore because there may be other dynamical processes which mix material at the photosphere itself as the accreting material hits the star we suspect that this approximation does not result in our neglecting an important bottleneck to mixing.

Combining equations~\eqref{eq:mlt} and~\eqref{eq:vc} we find
\begin{align}
	D \approx h \left(\frac{F_\star}{\rho}\right)^{1/3},
	\label{eq:conv}
\end{align}
which may be approximated with stellar scaling relations\footnote{See Appendix~\ref{appen:conv}.} as
\begin{align}
	D \approx 7\times 10^{13}\mathrm{cm^2 g^{-1}}\left(\frac{M}{M_\odot}\right)^{7/3} \left(\frac{\kappa}{10\mathrm{cm^2 g^{-1}}}\right)^{1/3}.
	\label{eq:convApprox}
\end{align}

\begin{figure*}
\includegraphics[width=1.0\textwidth]{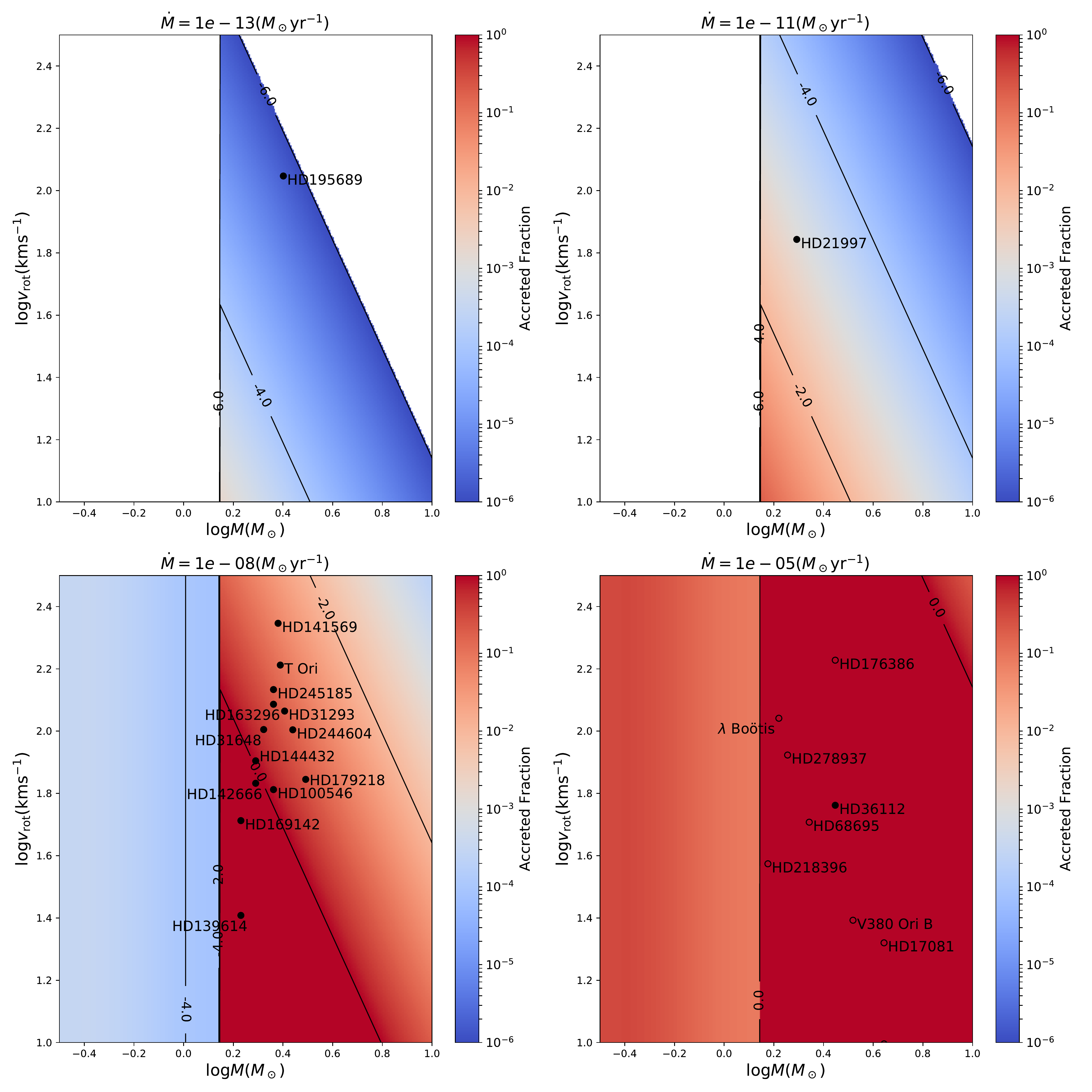}
\caption{The photospheric abundance of accreted material is shown in colour for main sequence stars as a function of stellar mass (horizontal) and rotation speed (vertical). Four accretion rates are shown: $10^{-13}M_\odot/\mathrm{yr}$ (top-left), $10^{-11}M_\odot/\mathrm{yr}$ (top-right), $10^{-8}M_\odot/\mathrm{yr}$ (bottom-left) and $10^{-5}M_\odot/\mathrm{yr}$ (bottom-right). Each star is placed on the panel which most closely matches its measured or predicted accretion rate. Stars with no such data are placed on the bottom-right panel and shown with open circles. Contours correspond to factor of $100$ increments. Regions with $f < 10^{-6}$ are shown in white. Note the sudden change around $1.4M_\odot$, corresponding to the onset of a surface convection zone which greatly enhances photospheric mixing. The accretion rate for HD~245815 is due to~\citet{2011AJ....141...46D}.}
\label{fig:summary}
\end{figure*}

\begin{figure}
\includegraphics[clip=,width=1.0\columnwidth]{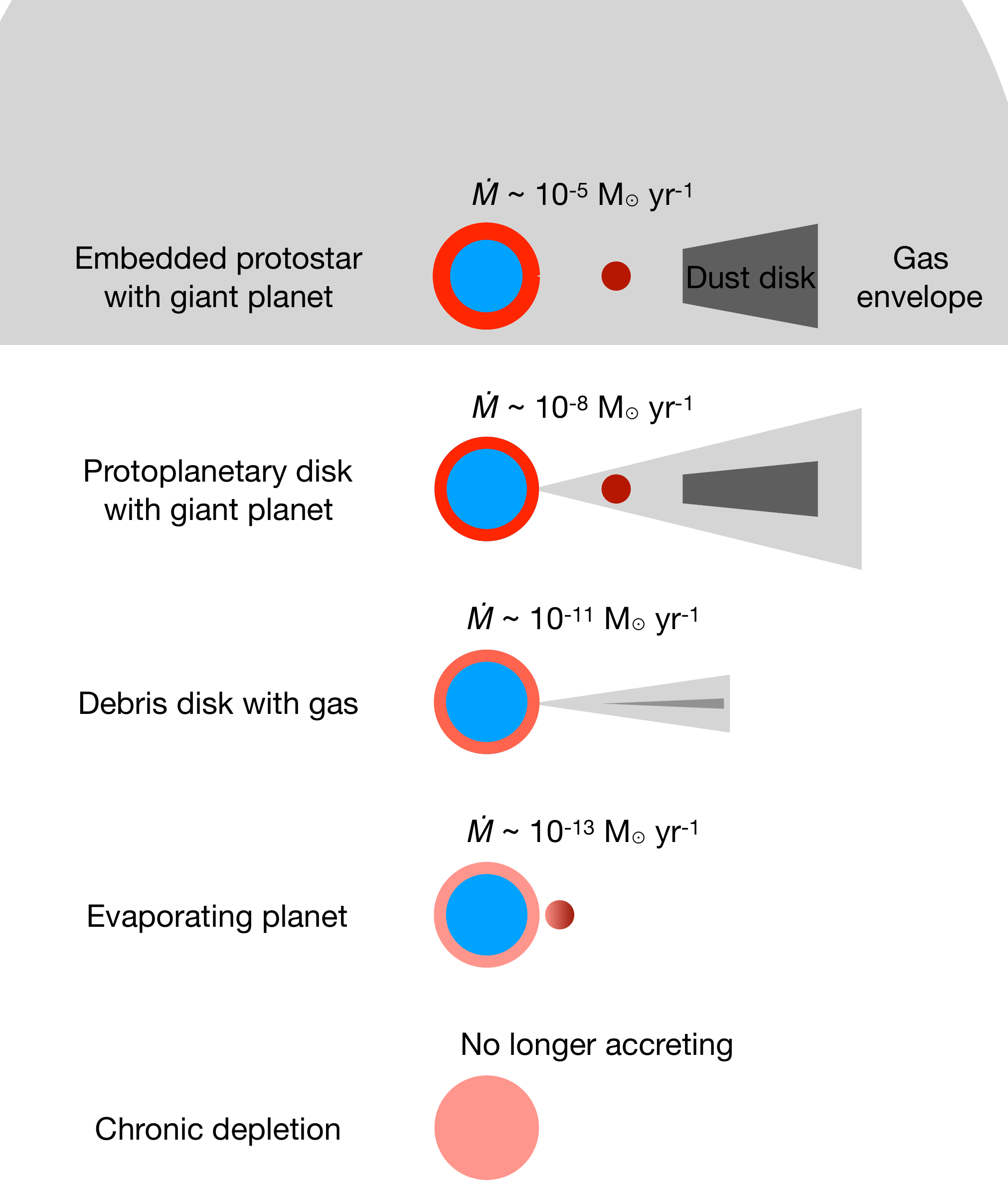}
\caption{Evolutionary phases where an early-type star may display surface abundance anomalies due to ongoing or recent accretion. Red material on the star indicates accreted material, while blue indicates primordial composition. Planet-induced radial dust depletion can provide selective accretion in the disk phase. Chronic depletion refers to hypothetical signatures which remain for $\gtrsim100\,$Myr after accretion ceases.}
\label{fig:sourcetypes}
\end{figure}

\section{Application to observations}

Sources of accretion onto stars include protoplanetary disks, debris disks, and evaporating or tidally disrupted planets or planetesimals. Intermediate-mass stars develop radiative envelopes fairly quickly, either on the Henyey track or immediately off the Hayashi track~\citep{doi:10.1063/1.4754323}. This allows the \texttt{CAM} framework to be applied to very young, disk-hosting early-type stars as well as main-sequence ones. Using known stellar parameters and the equations presented in the previous sections, one can calculate the fraction $f_{\rm ph}$ of accreted material in the stellar photosphere as a function of time and determine its composition. We detail the required equations and discuss various source types below.

We denote the observed fractional abundance of element X in the stellar photosphere with (X/H)$_{\rm obs}$, and use analogous notation for the reference stellar composition (``bulk'') and recently accreted material (``acc'').  The normalization with hydrogen (H) is arbitrary and just the most common convention.
The fraction of accreted material $f(r)$ is not a directly-observable quantity.
Rather, what is observed is the composition of the photosphere, (X/H)$_{\rm obs}$. For a given element X measured by a transition at wavelength $\lambda$, the inferred composition is
\begin{equation}
	\left(\frac{\rm X}{\rm H}\right)_{\rm obs} = \frac{\int_0^{R} \rho e^{-\tau_{\lambda}(z)} \left[f(z)\left(\frac{\rm X}{\rm H}\right)_{\rm acc} + (1-f(z))\left(\frac{\rm X}{\rm H}\right)_{\rm bulk}\right] dz}{\int_0^{R} \rho e^{-\tau_{\lambda}(z)} dz},
\end{equation}
where, $\tau_{\lambda}(z)$ is the optical depth at $z$ for light of wavelength $\lambda$, $\rm \left(X/H\right)_{acc}$ is the concentration of $X$ in the accreted material, and $\rm \left(X/H\right)_{\rm bulk}$ is the concentration of $X$ in the stellar material.
That is, the composition which is measured is a transmittance-weighted average over the star.

A reasonable approximation to this average is to identify the depth $z_{\mathrm{ph}}$ of the photosphere for a given spectral feature.
This depth is given by the implicit equation
\begin{equation}
	\tau_\lambda(z_\mathrm{ph}) = \int_{-\infty}^{z_\mathrm{ph}} \kappa_\lambda(\rho(z), p(z)) \rho dz \approx 1,
	\label{eq:cond}
\end{equation}
where $\kappa_\lambda(\rho, p)$ is the opacity of material at pressure $p$ and density $\rho$ at wavelength $\lambda$.
This expression in turn may be approximated as
\begin{align}
	\tau_\lambda(z_\mathrm{ph}) &= \int_{-\infty}^{z_\mathrm{ph}} \kappa_\lambda(\rho(z), p(z)) \rho dz\\
&\approx \int_{0}^{\Sigma(z_\mathrm{ph})} \kappa_\lambda(\rho(z), p(z)) d\Sigma\\
&\approx \int_{0}^{p(z_\mathrm{ph})} \kappa_\lambda(\rho(z), p(z)) g^{-1} dp\\
&\approx \kappa_\lambda(\rho_\mathrm{ph}, p_\mathrm{ph}) p_\mathrm{ph} g^{-1},
\end{align}
where we have made use of $d\Sigma = \rho dz$ and $p \approx \Sigma g^{-1}$ near the photosphere.
As $T$ is nearly constant throughout the photosphere, the equation of state determines $\rho$ from $p$ and so equation\ \eqref{eq:cond} may be written simply as
\begin{equation}
	\kappa_\lambda(p_\mathrm{ph}) p_\mathrm{ph} g^{-1} \approx 1.
	\label{eq:depth}
\end{equation}

To leading order the dependence on $\lambda$ may be dropped and $\kappa$ may be approximated by the Rosseland mean opacity, resulting in
\begin{equation}
	\kappa(p_\mathrm{ph}) p_\mathrm{ph} g^{-1} \approx 1.
	\label{eq:depth1}
\end{equation}
The solution to this equation yields a photospheric pressure, which may be converted to a depth using the equation of hydrostatic equilibrium $dp = -\rho g dz$.
The photospheric fraction $f_\mathrm{ph}$ is then given by equation\ \eqref{eq:fincrease}, which may be approximated as
\begin{align}
	f_\mathrm{ph} \approx \min\left(1,\frac{\dot{M} h}{4\pi R^2 \rho_\mathrm{ph} D_\mathrm{ph}}\right).
\end{align}
Note that we have used the form for cases in which $\rho^2 D$ increases with depth, as this is the only one relevant for the mixing mechanisms considered in the previous section.
A further approximation makes use of $h \ll R$ to write
\begin{align}
	M_{\mathrm{ph}} \approx 4\pi R^2 h \rho_{\mathrm{ph}},
\end{align}
where $M_{\mathrm{ph}}$ is the mass of the photosphere, which typically varies from $10^{-12} M_\odot$ for Sun-like stars to $10^{-10}M_\odot$ in more massive stars.
As such,
\begin{align}
	f_\mathrm{ph} \approx \min\left(1,\frac{\dot{M}}{M_{\mathrm{ph}}}\left(\frac{h^2}{D_\mathrm{ph}}\right)\right).
	\label{eq:fraction}
\end{align}
For context $h$ is typically of order $10^7\mathrm{cm}$ and $D_\mathrm{ph}$ varies between $10^{2}\mathrm{cm^2\,s^{-1}}$ when molecular diffusion dominates and $10^{14}\mathrm{cm^2\,s^{-1}}$ for stars with surface convection zones.
Finally, this fraction is related to the observed abundance of element X by the relation
\begin{equation}
	\left(\frac{\rm X}{\rm H}\right)_{\rm obs} = f\mathrm{ph}\left(\frac{\rm X}{\rm H}\right)_{\rm acc} + (1-f_{\mathrm{ph}})\left(\frac{\rm X}{\rm H}\right)_{\rm bulk}.
	\label{eq:abundance}
\end{equation}

The accreted fraction of the photosphere asymptotes rapidly to a constant except in the case where rotational mixing dominates, in which case there is a logarithmic correction in time. Figure~\ref{fig:summary} shows the fractional abundance of accreted material in the photosphere, neglecting the logarithmic correction, for main sequence stars as a function of mass and rotation speed for accretion rates $\dot{M}_{\rm acc}=10^{-5}$, $10^{-8}$, $10^{-11}$, and $10^{-13}M_\odot/\mathrm{yr}$. These rates represent accretion from the envelope of a protostar, a protoplanetary disk, a gas-rich debris disk, and an evaporating planet, respectively. Various evolutionary stages where a star may show surface abundance anomalies due to accretion are shown along with the relevant order-of-magnitude accretion rate in Figure~\ref{fig:sourcetypes}. We also include chronic (long-lived) depletion, which may occur if the total mass fraction accreted from material with an abundance anomaly is sufficient that a residual signature remains even after complete mixing, or if the mixing happens on a long timescale ($t_{\rm mix}\gtrsim 100\,$Myr).

The abundance of X in the mass reservoir providing the accretion is
\begin{equation}
{\rm \left(\frac{\rm X}{\rm H}\right)_{acc}} = \frac{1}{f}{\rm\left(\frac{\rm X}{\rm H}\right)_{obs}} - \frac{1-f}{f}{\rm \left(\frac{\rm X}{\rm H}\right)_{bulk}}
\label{eq:abundance2}
\end{equation}
Aside from calculating $f$, we need to assume a reference point for the bulk stellar abundances. Reasonable choices include the composition of the Sun \citep{Lodders2003, Asplundetal2009} or an average of nearby early-type stars \citep{Fossatietal2011, Martinetal2017}. In the current work, we favour the former for accuracy and precision, but the latter choice may be a better reference because it was obtained using spectroscopic methods nearly identical to those used in studies of protoplanetary disk hosting early-type stars \citep[e.g.][]{AckeWaelkens2004, Folsometal2012} and in our own ongoing observational efforts.

We now consider four source types where we can obtain new insight on the accreting material:
\begin{enumerate}
\item{Protoplanetary disks (e.g. HD~163296, HD~100546)}
\item{Debris disks (e.g. HD~141569~A, HD~21997)}
\item{Evaporating planets (e.g. HD~195689/KELT-9)}
\item{$\lambda\,$Bo\"{o}tis stars}
\end{enumerate}

\begin{figure}
\includegraphics[clip=,width=1.0\columnwidth]{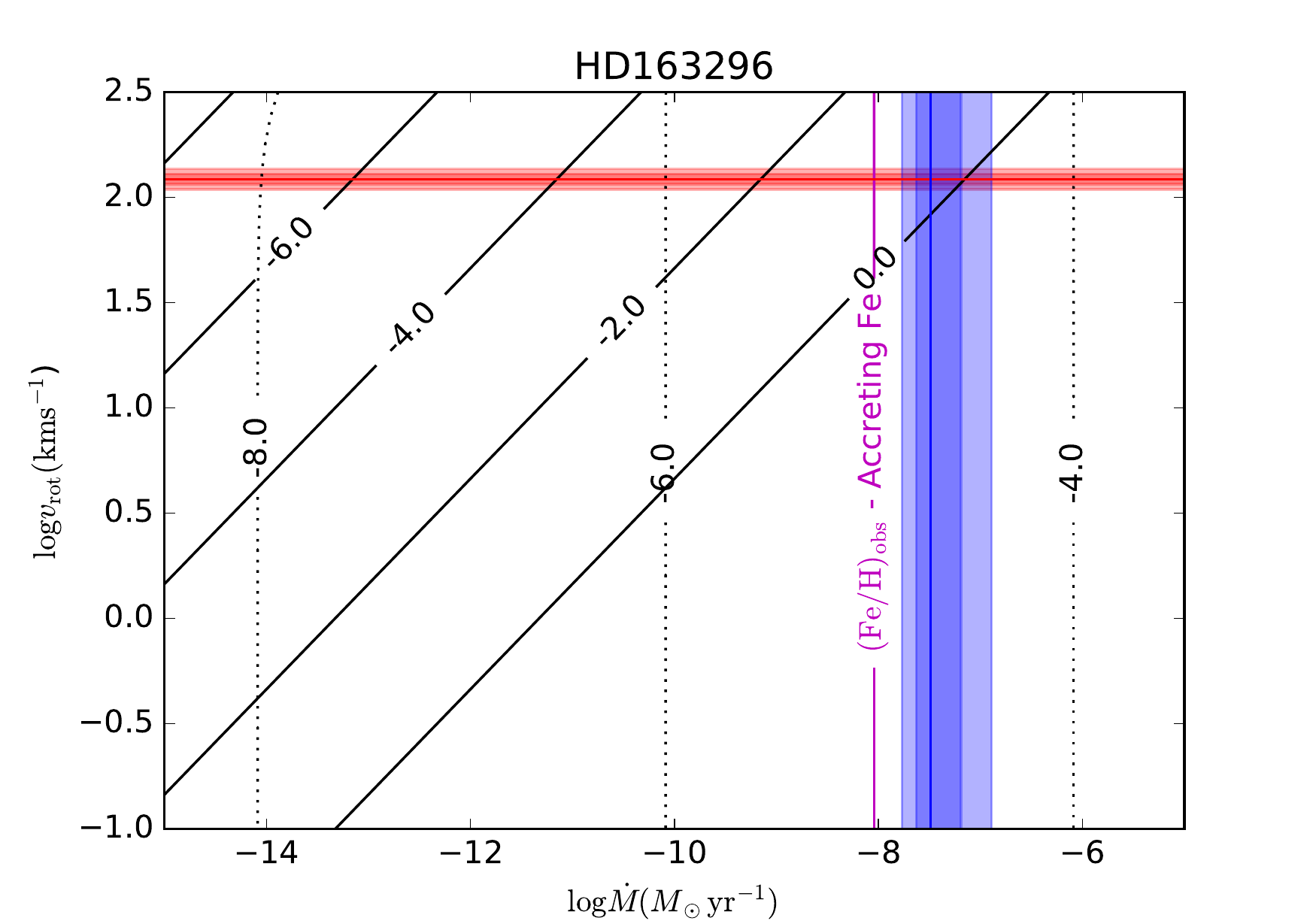}
\caption{HD~163296 hosts a full protoplanetary disk. We show the theoretical accreted photospheric mass fraction $f_{\rm ph}$ from accreting hydrogen-rich material (solid black lines) or accreting pure Fe (dotted), $f_{\rm ph}$ from the observed accretion and rotation rate (intersection of horizontal red and vertical blue bars, to be compared with the solid black lines), and an upper limit from the observed super-solar (Fe/H) of the star (vertical purple line, to be compared with the dotted black lines). The $\pm \sigma$ and $\pm 2\sigma$ contours for $\log v_\mathrm{rot}$ are shown in red while those for $\log \dot{M}_{\rm acc}$ are shown in blue.}
\label{fig:hd163296}
\end{figure}

\begin{figure}
\includegraphics[clip=,width=1.0\columnwidth]{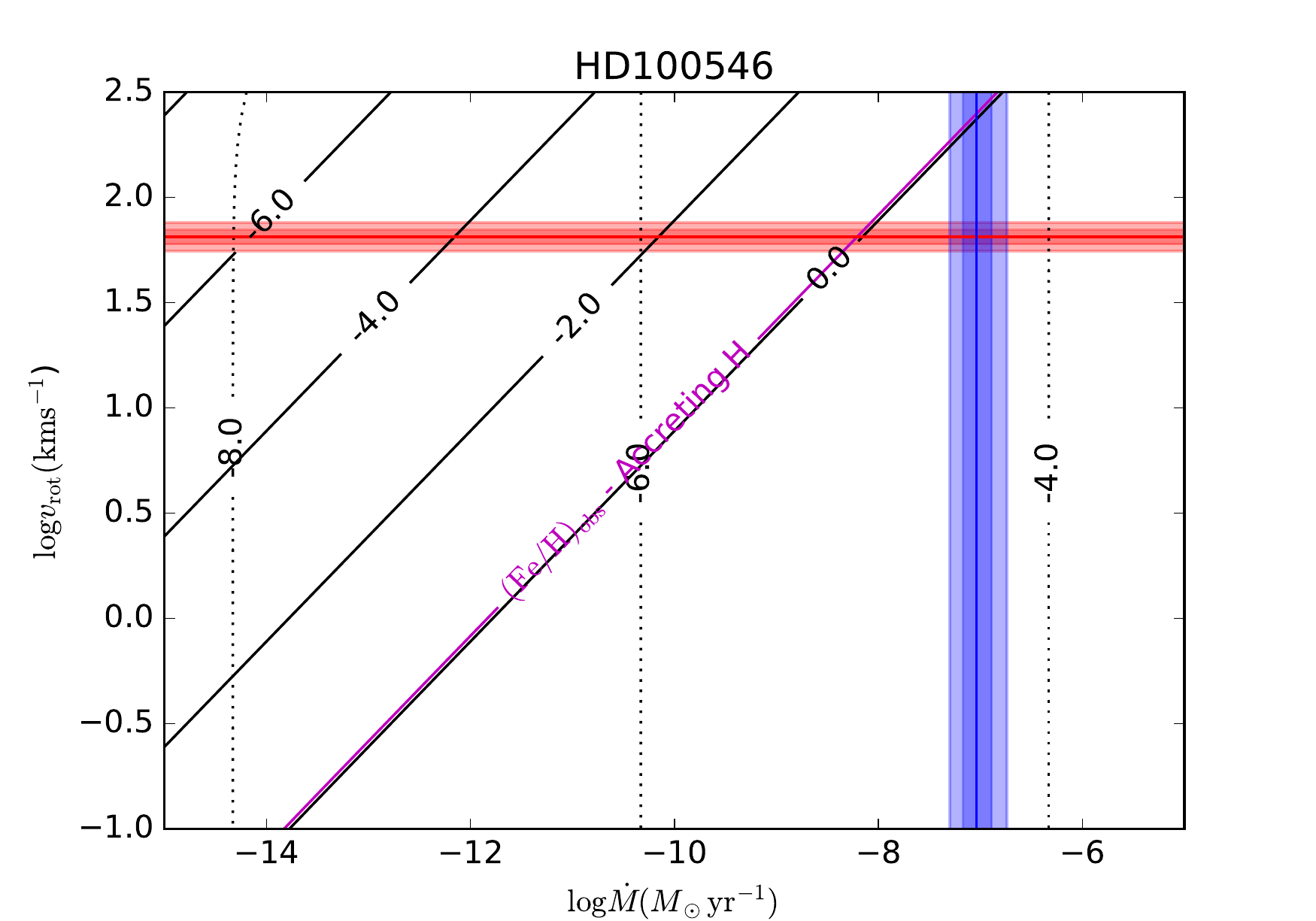}
\caption{HD~100546 hosts a transitional protoplanetary disk. We show the theoretical accreted photospheric mass fraction $f_{\rm ph}$ from accreting hydrogen-rich material (solid black lines) or accreting pure Fe (dotted, not relevant in this source), $f_{\rm ph}$ from the observed accretion and rotation rate (intersection of horizontal red and vertical blue bars, to be compared with the solid black lines), and a lower limit from the observed low (Fe/H) of the star (diagonal purple line, to be compared with the solid black lines and the red and blue bars). The $\pm \sigma$ and $\pm 2\sigma$ contours for $\log v_\mathrm{rot}$ are shown in red while those for $\log \dot{M}_{\rm acc}$ are shown in blue.}
\label{fig:hd100546}
\end{figure}

\begin{figure}
\includegraphics[clip=,width=1.0\columnwidth]{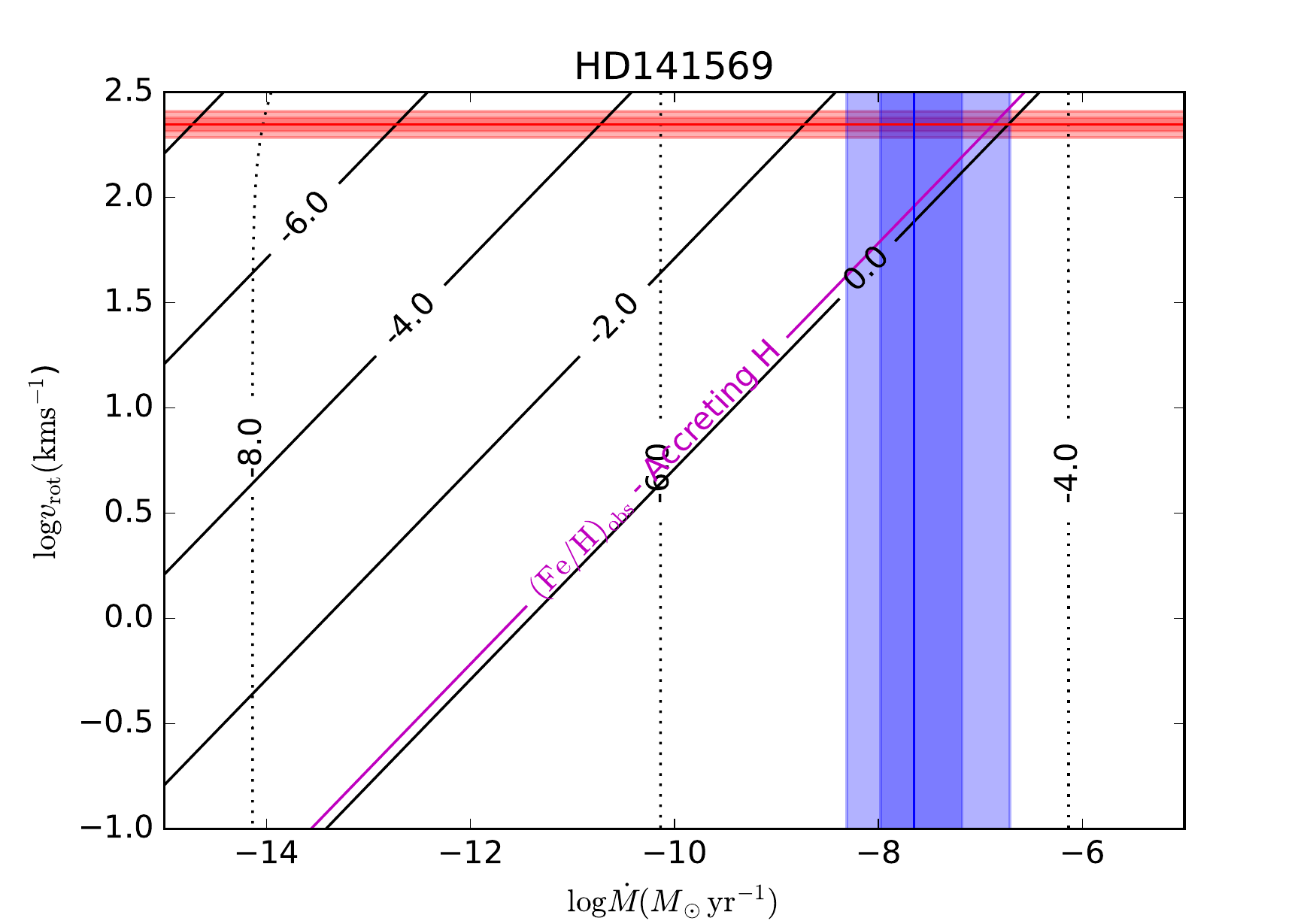}
\caption{HD~141569 hosts a hybrid disk. We show the theoretical accreted photospheric mass fraction $f_{\rm ph}$ from accreting hydrogen-rich material (solid black lines) or accreting pure Fe (dotted, not relevant in this source), $f_{\rm ph}$ from the observed accretion and rotation rate (intersection of horizontal red and vertical blue bars, to be compared with the solid black lines), and a lower limit from the observed low (Fe/H) of the star (diagonal purple line, to be compared with the solid black lines and the red and blue bars). The $\pm \sigma$ and $\pm 2\sigma$ contours for $\log v_\mathrm{rot}$ are shown in red while those for $\log{(\dot{M}_{\rm acc})}$ are shown in blue.}
\label{fig:hd141569}
\end{figure}

\begin{figure}
\includegraphics[clip=,width=1.0\columnwidth]{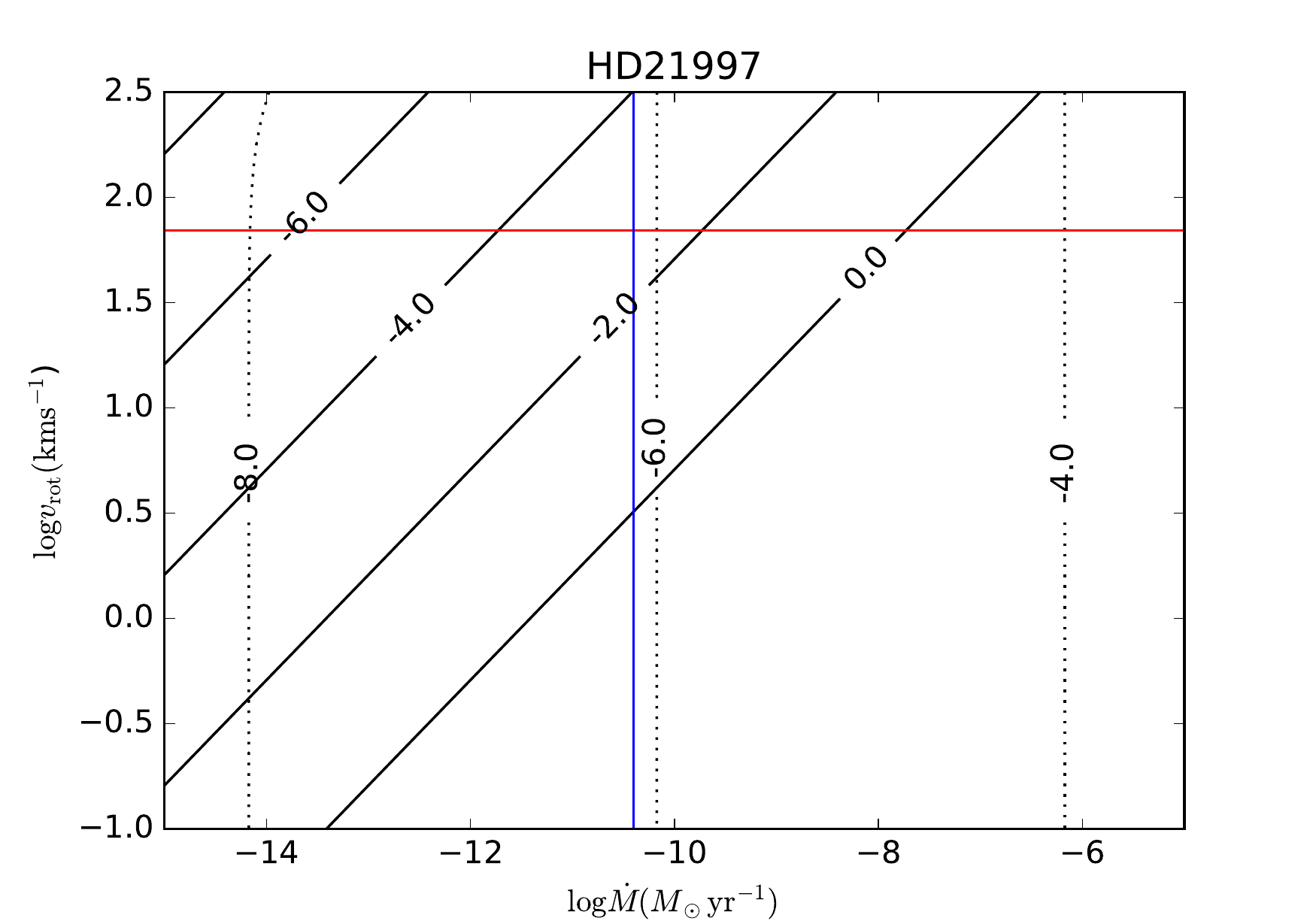}
\caption{HD~21997 hosts a hybrid disk. The high $v_{\rm rot}$, and the accretion rate predicted by the \citet{Kraletal2016, Kraletal2017} model, $\dot{M}_{\rm acc}\sim10^{-11}\,$M$_{\odot}\,$yr$^{-1}$, yields $f_{\rm ph}= 4\times 10^{-3}$. We show the theoretical accreted photospheric mass fraction $f_{\rm ph}$ from accreting hydrogen-rich material (solid black lines) or accreting pure Fe (dotted), $f_{\rm ph}$ from the model-predicted accretion and observed rotation rate (intersection of horizontal red and vertical blue bars). Note that no uncertainties are given for $v_{\rm rot}$ because none were reported from observations, and likewise none are given for $\dot{M}_{\rm acc}$ because none were given in the predicted rates. If the accreted material is largely C and O, the stellar photospheric abundances of these elements may be dominated by outgassed debris disk material.}
\label{fig:hd21997}
\end{figure}

We accompany each of the example stars with a figure; Figs.~\ref{fig:hd163296}, \ref{fig:hd100546}, \ref{fig:hd141569}, \ref{fig:hd21997}, and \ref{fig:hd195689} respectively. In addition we discuss polluted white dwarfs, which provide an additional set of constraints on similar systems at a later stage of evolution. Input and calculated parameter values are summarized in Table~\ref{tab:sources}. The figures show the newly accreted photospheric mass fraction, $f_{\rm ph}$, calculated from Eq.~\ref{eq:fraction} (solid black lines). As a comparison with this theoretical calculation, $f_{\rm ph}$ can also be estimated from observations using the measured stellar rotation rate, $v_{\rm rot}$, and the measured or predicted accretion rate, $\dot{M}_{\rm acc}$ (intersection of red horizontal and blue vertical lines). Observed $v_{\rm rot}$ values are shown for each star, while $\dot{M}_{\rm acc}$ is either observational, or predicted from models or theory. 

Where available, e.g. for HD~100546 (Fig.~\ref{fig:hd100546}), the observed (Fe/H)$_{\rm obs}$ is used to calculate a lower limit on $f_{\rm ph}$ (purple diagonal line), which corresponds to the fraction of the photosphere that needs to be replaced with pure H in order to decrease (Fe/H) to the observed level. If (Fe/H)$_{\rm obs}$ is super-solar, e.g. for HD~163296, we obtain instead an upper limit on $f_{\rm ph}$ (purple vertical line) under the assumption that the accretion flow contains only Fe. An analogous reasoning would apply to any other element, here we use iron because it is easy to determine and indicative of a depletion of dust from the accreted material. Specific elemental signatures may be different in specific cases, depending on the origin of the accreted material.


\begin{table*}
\centering
\begin{tabular}{ l c c c c c c c c l }
\hline\hline
Star	& Type & $\log\left(\frac{\rm Fe}{\rm H}\right)_{\rm obs}$ & $v_{\rm rot}$ & Age & $\dot{M}_{\rm acc}$ & $\log t_\mathrm{ph}$ & $\log f$& $\log f$ & Notes \\ 
& & & $\left({\rm km}\,{\rm s}^{-1}\right)$ & (Myr) & $\left(M_{\odot}\,{\rm yr}^{-1}\right)$ & ($s$) & (Pred.) & (Pred. Heavy) & \\
\hline
HD~163296 & A3 & $-4.39^{+0.15}_{-0.15}$ & $122.0\pm 3.0 $ & $7.56^{+ 2.17 }_{-2.17}$ & $-7.49^{+ 0.3 }_{-0.14} $ & $3.9$&$-0.33$ & $-4.70$ & \begin{minipage}[t]{0.37\columnwidth}\citet{Folsometal2012,Fairlambetal2015}\end{minipage}\\
HD~100546 & A0 & $-5.67^{+0.08}_{-0.08}$ & $64.9\pm 2.2 $ & $7.02^{+ 1.49 }_{-1.49}$ & $-7.04^{+ 0.15 }_{-0.13} $ & $3.5$&$+0.00$ & $-4.45$ & \begin{minipage}[t]{0.37\columnwidth}\citet{Folsometal2012,Fairlambetal2015,Kamaetal2016b}\end{minipage}\\
HD~141569 & B9.5 & $-5.25^{+0.32}_{-0.32}$ & $222.0\pm 7.0 $ & $9.0^{+ 4.5 }_{-4.5}$ & $-7.65^{+ 0.47 }_{-0.33} $ & $3.7$&$-0.92$ & $-4.76$ & \begin{minipage}[t]{0.37\columnwidth}\citet{Folsometal2012,Fairlambetal2015}\end{minipage}\\
HD~21997 & A3 & -4.47$^{\star}$ & $69.7^{\star}$ & $30.0$ & $-10.4^{\dagger}$ & $4.3$&$-2.40$ & $-5.98$ & \begin{minipage}[t]{0.37\columnwidth}\citet{Viganetal2012,Kraletal2016,Kraletal2017,Netopiletal2017,Private}\end{minipage}\\
HD~195689 & A0 & $-4.53^{+0.20}_{-0.20}$ & $111.4\pm 1.3 $ & $300.0$ & $-13.0^{+ 2.0 }_{-2.0}$$^{\dagger}$ & $3.6$&$-5.98$ & $-7.52$ & \begin{minipage}[t]{0.37\columnwidth}\citet{Gaudietal2017}\end{minipage}\\
\hline
\end{tabular}
\caption{Stars chosen to demonstrate the application of \texttt{CAM} to study the composition of accreting material. $t_{\mathrm{ph}}$ is $h^2/D$ evaluated at the photosphere. The predicted $\log f$ was calculated using equation~\eqref{eq:fincrease}, with molecular weight gradients incorporated in the heavy case on the assumption of Fe accretion.~$^{\star}$ -- (Fe/H)$_{\rm obs}$ for HD~21997 was determined photometrically and no errorbar was reported. Likewise no errorbar was reported for $v_\mathrm{rot}$ for this star. $^{\dagger}$ -- Accretion rates predicted from models.}
\label{tab:sources}
\end{table*}

\subsection{Protoplanetary disks}\label{sec:haebe}

Three immediate applications of our method suggest themselves for young B-/A-/F-type stars hosting protoplanetary disks: (i) measuring the gas-to-dust mass ratio, $\Delta_{\rm g/d}$; (ii) measuring the fraction of each element locked in refractory particles; and (iii) assessing whether or not a given young star has developed a radiative envelope. We discuss each of these in turn.

\begin{enumerate}
\item{\emph{An independent measurement of the gas-to-dust mass ratio in the inner disk.}

For $1$--$10\,$Myr old B to mid-F type stars hosting protoplanetary disks a correlation has been found between the presence of dust-depleted gaps in the disk and a decreased abundance of refractory elements on the stellar photosphere \citep{Kamaetal2015}. That study found the photospheric refractory abundances to correspond to $\Delta_{\rm g/d}$ values as high as $1000$, but did not account for the cases where $f<1$ (i.e. where accretion does not replace the entire photosphere in steady state). Applying $f$ values calculated with the formalism developed in the current paper, we expect to find some $\Delta_{\rm g/d}$ values much larger than before, and in even better agreement with the dust depletions directly determined from spatially resolved continuum observations~\citep[e.g.][but a limitation of that data is that the gas surface density is measured indirectly via CO]{vanderMareletal2015, vanderMareletal2016}.

Two cases in point are HD~163296 (Figure~\ref{fig:hd163296}) and HD~100546 (Figure~\ref{fig:hd100546}), stars hosting a full and a transitional disk, respectively. For both of these stars, the high theoretical $f_{\rm ph, HD~163296}=0.47$ and $f_{\rm ph, HD~100546} = 1.0$ (Table~\ref{tab:sources}) confirm that the photospheric abundance pattern closely follows the abundance pattern of freshly accreted material.

HD~163296 is accretion-dominated but has slightly super-solar abundances of refractory elements \citep{Folsometal2012} and so $\rm (Fe/H)_{acc}\gtrsim(Fe/H)_{\odot}$. This implies the relevant features to examine in Figure~\ref{fig:hd163296} are the theoretical $f_{\rm ph}$ from pure-Fe accretion (dotted black lines) and the upper limit on $f_{\rm ph}$ from the observed photospheric abundance (Fe/H)$_{\rm obs}$. An enhancement of refractories is consistent with disk models which suggest the disk material around HD~163296 is gas-poor \citep[$\Delta_{\rm g/d}\approx 20$,][]{Bonebergetal2016}.

For HD~100546, the observed abundance of refractory elements is low, (Fe/H)$_{\rm obs}=-5.67$ \citep{Kamaetal2016b}. Following the theoretical calculations for H-rich accretion in Figure~\ref{fig:hd100546} (solid black lines) and the observed rotation and accretion rate constraints (red and blue bars) gives $f_{\rm ph}=1.0$ and $\rm (Fe/H)_{acc}=(Fe/H)_{obs}$. The observed abundance is a factor of $10$ below solar and gives $\Delta_{\rm g/d}=1000$, consistent with the structure of the inner dust cavity in gas-dust disk models \citep{Brudereretal2012, Kamaetal2016b}.}
\item{\emph{A quantification of the refractory \emph{vs} volatile fraction of any element measured in the photospheric spectrum.}

As noted above, the most refractory elements are, on average, depleted in abundance by $0.5\,$dex in the photospheres of stars with a transitional disk \citep{Kamaetal2015}.
Such disks have extended radial zones where the surface density of large particles is reduced by up to several orders of magnitude.
In contrast, the same study found the most volatile elements, such as C and O, have similar abundances for all disk-hosting early-type stars regardless of the disk structure.
For elements of intermediate sublimation temperature, however, the depletion may be only partial, and this offers a way of measuring the refractory fraction of a given element.
A key example which we will address in a follow-up paper is sulfur, which is predominantly refractory in primitive solar system meteorites, but whose refractory fraction in protoplanetary disks has not yet been observationally determined.}

\item{\emph{An independent assessment of whether or not a star has developed a radiative envelope.}

For early-type protoplanetary disk hosts, it is usually possible to measure the stellar photospheric abundances \citep{AckeWaelkens2004, Folsometal2012} \emph{and} the accretion rate onto the star \citep[e.g.][]{Mendigutiaetal2011, Fairlambetal2015}. This is important for B/A/F-type stars, in particular, because early on such stars are expected to be fully convective~\citep{1961PASJ...13..450H}, but once on the main sequence their upper atmospheres are radiative~\citep{1992isa..book.....B}. This entails a dramatic shift in the diffusivity near the photosphere, and so the change to a radiative envelope ought to be detectable even in highly uncertain measurements, if the accreting material has a composition substantially depleted in refractory elements or has an otherwise significantly different composition from the bulk stellar material. Whether or not a young star is convective can be a key indicator of its age, which is normally quite difficult to determine~\citep{1999MNRAS.310..360T, Soderblometal2014}.

To understand this quantitatively, recall that the fraction of accreted material in the photosphere is given by equation\ \eqref{eq:fraction} as
\begin{align}
	f_\mathrm{ph} \approx \frac{\dot{M}}{M_{\mathrm{ph}}}\left(\frac{h^2}{D_\mathrm{ph}}\right).
\end{align}
For a species where the stellar and accreted material have very different abundances this fraction is readily measured and closely related to any depletion/enhancement by equation\ \eqref{eq:abundance}.
Inserting equation\ \eqref{eq:convApprox} we find
\begin{align}
	f_\mathrm{ph} \approx 10\mathrm{s}\frac{\dot{M}}{M_{\mathrm{ph}}}\left(\frac{M}{M_\odot}\right)^{5/12} \left(\frac{\kappa}{10\mathrm{cm^2 g^{-1}}}\right)^{-1/3}.
	\label{eq:astars}
\end{align}
The upshot of this being so much less than any of the mechanisms active in radiative stars is that it provides a robust means of determining which stars have formed their radiative envelopes and which ones are still fully convective.

If $f_\mathrm{ph}$ can be measured or bounded and is found to be much greater than Eq.\ \eqref{eq:astars} suggests -- such as would be inferred if a significant depletion of refractories is found -- mixing must be much less efficient than convection and so the upper regions of the star must be radiative. Conversely, if $f_\mathrm{ph}$ is found to be consistent with this equation then this strongly suggests a convection zone. A strong case for this might be made if a disk is robustly established to have a significant dust depletion in an inner cavity, but the central star is observed to have a high abundance of refractory elements. 

The power of this test in the face of the typically order of magnitude of uncertainty in the accretion rates and up to a factor of two in abundances comes from the massive difference in mixing efficiency between radiative mechanisms and bulk turbulent convection.}
\end{enumerate}

\subsection{Debris disks}\label{sec:debrisdisks}

In debris disks, the gas-to-dust mass ratio is generally low, $\Delta_{\rm g/d}\lesssim 1$, and the dust dynamics are independent of the gas. A slew of recent results has revealed that significant quantities of carbon and oxygen gas can be present, either in the form of CO molecules, or neutral or ionized C or O atoms. The debris disks with gas are found around early-type (A, F) stars and fall into two categories: those for which the C and O gas mass can be explained with outgassing of CO and H$_{2}$O from exocomets, such as $\beta\,$Pic, HD~181327, and Fomalhaut; and the so-called ``hybrid'' disks such as HD~141569~A and HD~21997, which have a large gas mass and which may be primordial, gas-rich disks in a late stage of dissipation \citep{Mooretal2011, Pericaudetal2016, doi:10.1093/mnras/stw1216, Kraletal2017, Hughesetal2017, Matraetal2017a, Matraetal2017b, Mooretal2017}. A major difference between the two types is that the hybrid disks are likely still dominated by H$_{2}$, while in the outgassing disks C, O, and H all likely have an order-of-magnitude similar number density because the hydrogen comes from H$_{2}$O photodissociation.

Models of exocometary gas production and viscous disk evolution allow predictions to be made for the accretion rate of C- and O- rich gas onto the host stars debris disks, and these typically lie in the range of $10^{-13}$ to $10^{-11}\,$M$_{\odot}\,$yr$^{-1}$ \citep[][and private communication]{Kraletal2016, Kraletal2017}. From $\dot{M}_{\rm acc}$, the accreted fraction $f$ of total photosphere mass can be calculated.

HD~141569~A is a heavily accreting hybrid disk system. As shown in Fig.~\ref{fig:hd141569}, from the apparently extremely high accretion rate, we predict $f_{\rm ph}=0.12$, in spite of the very high $v_{\rm rot}\,\sin{(i)}$. In comparison, the observed photospheric Fe/H abundance gives a lower limit of $f_{\rm obs}\geq 0.9$. Considering the errorbars on $\dot{M}_{\rm acc}$ and $v_{\rm rot}$, the theoretical and observationally constrained $f$ are consistent within a factor of a few and suggest the photosphere consists entirely, or almost entirely, of freshly accreted, refractory-poor material. This implies that the bulk of the measured accretion is indeed due to H and other volatile elements, indicating that at least the inner disk is relatively gas-rich, and indeed a large-scale CO gas disk has been resolved with ALMA~\citep{2016ApJ...818...97F,2016ApJ...829....6W}.

HD~21997, a hybrid disk shown in Fig.~\ref{fig:hd21997}, has no reported detection of accretion onto the star. Furthermore, a precise, spectroscopic (Fe/H)$_{\rm obs}$ was not immediately available from the literature. We use an $\dot{M}_{\rm acc}$ value from the \citet{Kraletal2016} model to obtain $f_{\rm ph}=4\times 10^{-3}$. This accreted mass fraction is very low in absolute terms. It is thought to be a hybrid disk which may have retained a relatively large amount of primordial H$_2$ from the protoplanetary disk stage. The photospheric abundances of the star offer a way to distinguish a primordial origin of the gas from the cometary outgassing scenario. In the latter case, the volatile elements C and O would be accreted in large quantities with barely any hydrogen, while in the primordial case the accretion would be hydrogen-dominated. Thus measuring the stellar C and O abundances can serve as a test of models of gas-rich debris disks.

\begin{figure}
\includegraphics[clip=,width=1.0\columnwidth]{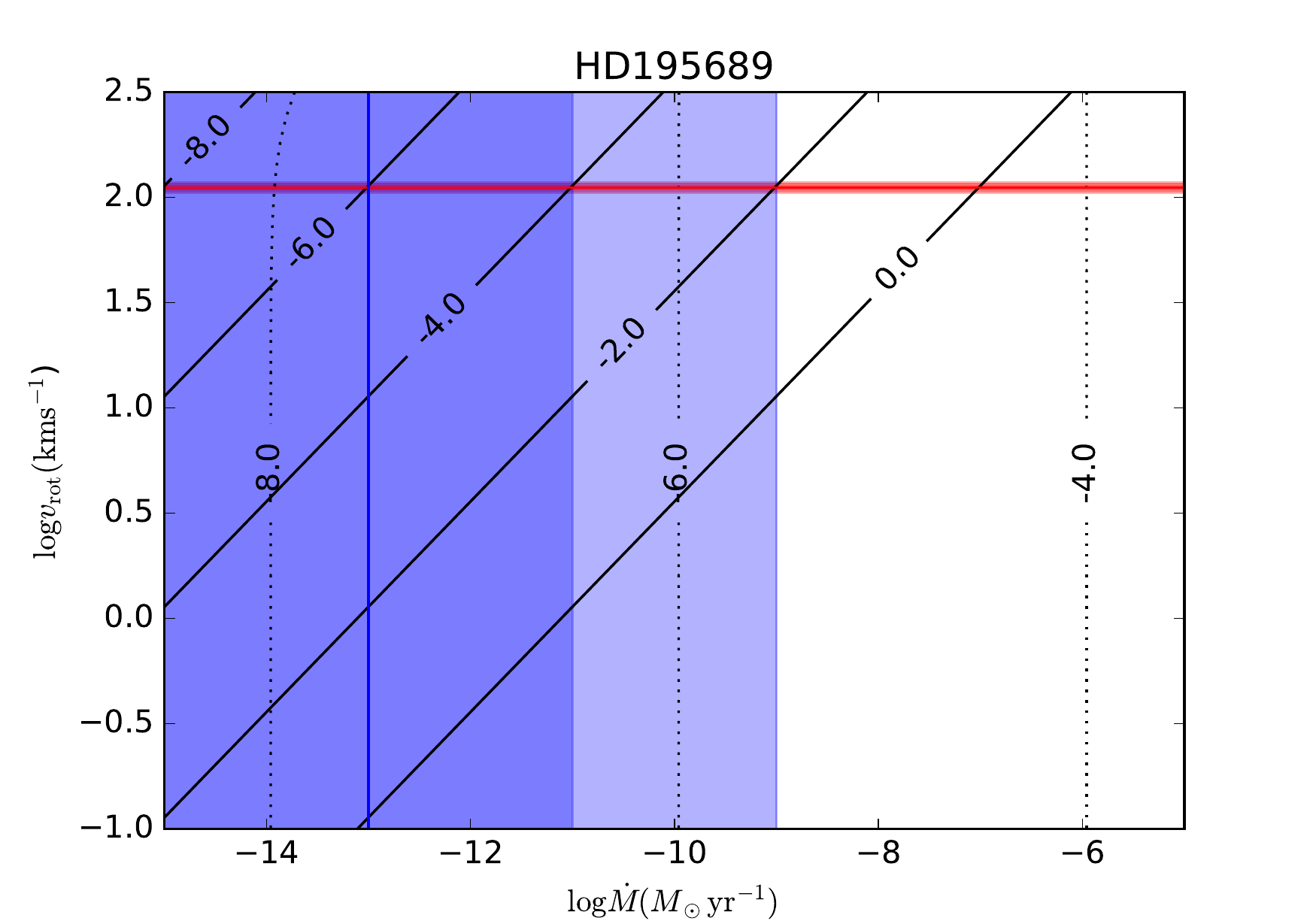}
\caption{HD~195689 hosts the hottest known Hot Jupiter, with $T_{\rm eq}=4600\,$K \citep[KELT-9b,][]{Gaudietal2017}. We show the theoretical accreted photospheric mass fraction $f_{\rm ph}$ from accreting hydrogen-rich material (solid black lines) or accreting pure Fe (dotted), and $f_{\rm ph}$ from the model-predicted accretion and observed rotation rate (intersection of horizontal red and vertical blue bars). The $\pm \sigma$ and $\pm 2\sigma$ contours for $\log v_\mathrm{rot}$ are shown in red while those for $\log{(\dot{M}_{\rm acc})}$ are shown in blue.}
\label{fig:hd195689}
\end{figure}

\subsection{Evaporating or disrupted planets}\label{sec:planets}

A planet sufficiently close to its star will undergo gradual evaporation or even total disruption. Some or most of the ejected material will accrete onto the star, where it may potentially be observed as photospheric contamination. This  possibility was first suggested for the diffusively mixing early-type field stars by \citet{Jura2015} in an effort to explain why stars without any apparent circumstellar mass reservoir display the \lboo\ abundance anomaly. 

The ultra-violet (UV) heating flux intercepted by a planet returns through two cooling channels -- a radiative component and an escaping particle flux component \citep{MurrayClayetal2009}. The escaping particle flux at high UV flux ($F_{\rm UV}>F_{\rm 0}$, where $F_{\rm 0}=450\,{\rm erg\,cm^{-2}\,s^{-1}}$) is given by
\begin{equation}
\dot{M}_{\rm UV} \approx 4\times 10^{12} \left(\frac{F_{\rm UV}}{F_{\rm 0}}\right)^{1/2}\,\left[\frac{\rm g}{\rm s}\right]
\label{eq:dotMUV}
\end{equation}

The most heavily irradiated exoplanet known to-date is HD~195689b (KELT-9b), the first planet found around a B9/A0 star \citep{Gaudietal2017}. Using Eq.~\ref{eq:dotMUV} and adopting $L_{\rm UV}\approx 10^{-3}\,$L$_{\odot}$ for the UV-luminosity of HD~195689, we find $\dot{M}_{\rm UV}=5.7\times 10^{12}\,$g$\,$s$^{-1}$ or $10^{-13}\,$M$_{\odot}\,$yr$^{-1}$. The photospheric mass fraction of accreted planetary material is shown in Fig.~\ref{fig:hd195689} for $\log_{10}(\dot{M}_{\rm acc}/(M_{\odot}yr^{-1}))=-13\pm2$.

Taking the mean of the logarithmic range of possible mass loss rates from the planet, we find an accreted photospheric mass fraction of $f_{\rm ph}=10^{-6}$.  For the highest potential planetary evaporation rate, $M_{\rm acc}=10^{-11}\,$M$_{\odot}\,$yr$^{-1}$, we find $f_{\rm ph}\approx 3\times10^{-5}$. If, however, this mass fraction consists primarily of non-hydrogen elements, it may still lead to a detectable enhancement in their stellar photospheric abundance.

HD~195689b and HD~185603b \citep[MASCARA-2b and KELT-20b,][]{Talensetal2017a, Lundetal2017} are the only Hot Jupiter planets known to-date to orbit main sequence A0 stars. Continuation of programmes such as KELT and MASCARA is expected to increase the sample size of heavily irradiated exoplanets with a high mass loss rate. Applying \texttt{CAM} to their host stars can potentially provide new information on the mass loss rate and elemental abundance ratios of such planets. In particular, it may be feasible to obtain a completely independent measurement of the C/O ratio in the atmosphere of giant planets.

Earth-like or Super-Earth planets undergoing disruption or active evaporation may also provide a sufficient effective accretion rate to yield detectable signatures. Accreting a $1\,$M$_{\oplus}$ planet over $1\,$Myr yields an average accretion rate $\dot{M}_{\rm acc}=10^{-12}\,$M$_{\odot}\,$yr$^{-1}$. As the accreted material would consist mostly of refractory elements, the relative enhancement of each such element could potentially be of order a factor of a few. Such a signature would remain visible in the photosphere for $0.1$ to $1\,$Myr. 

It has been suggested that elemental abundances in sun-like stars with planets are modified by a large fraction of refractories being locked in rocky planets or by the accretion of entire giant planets \citep[e.g.][]{Melendezetal2009, Ramirezetal2009, Wilsonetal2017}. However, those analyses have postulated an effect on the bulk composition of the star, or on its convection zone which is substantial for low-mass stars. Other studies suggest a relation between the overall metallicity of the system and the masses of its planets \citep{Buchhaveetal2014}. Our method offers a snapshot in time of the composition of material currently or recently ($\leq1\,$Myr) lost from a planet and accreted onto its host star.

\subsection{$\lambda\,$Bo\"{o}tis Stars}

Significant photospheric depletion of refractory elements, called the \lboo\ phenomenon, is seen in $\sim2\,$\% of B to mid-F type stars \citep{PaunzenGray1997, Paunzenetal2001, Paunzen2001, Murphyetal2015}. The \lboo\ abundance pattern is seen in young, disk-hosting stars with ages $\sim0.1$ to $10\,$Myr (Section~\ref{sec:haebe}) and in stars as old as $1\,$Gyr, on the order of their main--sequence lifetimes~\citep{2017MNRAS.466..546M}. Some \lboo\ stars possess debris disks \citep{Draperetal2016}, but the fraction of confirmed debris disks was found to be statistically indistinguishable from that in non-\lboo\ sources \citep{Grayetal2017}. Although the fraction of stars with the \lboo\ phenomenon among protoplanetary disk hosts \citep[$\gtrsim30\,$\%,][]{Folsometal2012} is much higher than among the total B to mid-F sample, the refractory element depletion in any given star does not correlate with its age \citep{IlievBarzova1995}. Firstly, this suggests that most young \lboo\ stars can indeed be explained as temporary beacons of the selective accretion of gas rather than dust \citep{VennLambert1990}, and indeed it has been shown that in young stars the \lboo\ phenomenon is a result of the star hosting a dust-depleted ``transitional'' disk \citep{Kamaetal2015}. Secondly, a small fraction of \lboo\ stars appear to survive to old age with a long-lasting or chronic depletion. A possible explanation is that the transitional disk phase for such \lboo\ stars was early and long-lasting enough for the dust-poor material to compose a significant fraction of the total mass of the star.

Two alternative explanations emerge in the context of this work.
First, it is possible that the transitional disk phase lasted for a substantial fraction of the age of the star but did not result in the accretion of a substantial fraction of its mass.
In that case the accretion signatures would still be apparent in its photosphere according to equation~\eqref{eq:sigbt} even though the accretion might have ceased long ago.
Secondly, and potentially in better agreement with the fraction of stars observed to have such accretion, it is possible that the transitional disk phase was both short relative to the age of the star and provided comparatively little material, but that the diffusivity in the star fell significantly towards the end of the accretion period.
In that case equation~\eqref{eq:dnew} indicates that the time over which the accreted material should be observable in the photosphere ought to be enhanced by a factor of $D_{\mathrm{old}} / D_{\mathrm{new}}$.

Under this second story, long-term accretion results in material mixing to significant depths, as $\Sigma_0$ may become quite large.
At such depths the dominant mechanism for mixing would be rotational mixing (e.g. meridional flows).
If the star magnetically breaks~\footnote{When the star is young it could well have both winds and a fossil magnetic field.} from a rotation rate of $\Omega_{\mathrm{old}}$ to one of $\Omega_{\mathrm{new}}$ towards the end of the accretion period then the timescale over which the accreted material ought to be observable is of order
\begin{align}
	t_\mathrm{obs} \approx t_{\mathrm{acc}} \left(\frac{\Omega_{\mathrm{old}}}{\Omega_{\mathrm{new}}}\right)^2.
\end{align}
This possibility is particularly interesting because during the accretion process the disk may transfer significant quantities of angular momentum to the star, even bringing it to near breakup with relatively modest amounts of material\footnote{On the order of $0.05M_\odot$~\citep{refId0LBL} would suffice.}, and so it is quite natural for the star to spin up during the transitional disk phase and spins down shortly afterwards.
If the spin-down occurs on the same time-scale as the accretion or faster and if the spin-down is significant then the historical accretion could remain observable for considerably longer than would otherwise be expected.
This is likely not long enough for the prototypical \lboo\ but with either a somewhat longer accretion phase or a more severe spin-down may explain at least some of the observed systems.

\subsection{White Dwarf Constraints}\label{sec:wd}

After going through an Asymptotic Giant Branch (AGB) phase, the stars in many of the systems we have considered will become white dwarfs.
These commonly host planetismal debris \citep{Jura2003, Reachetal2005, Zuckermanetal2010, Farihietal2009, Koesteretal2014, Juraetal2009} and may also host planets and planetary remnants \citep{Kleinetal2010}.
In such cases not only is accretion onto the white dwarf a probe of the current circumstellar elemental composition \citep{Zuckermanetal2007, Gaensickeetal2012, Farihietal2013, Xuetal2014} but also potentially of earlier chemical processing.
It is possible that such signatures, when combined with models of planetesimal formation and evolution, may provide further insights into the earlier volatile/refractory partitioning of elements in these systems \citep{Kleinetal2010, JuraYoung2014}.

\section{Conclusions}
We present a theoretical framework (Contamination by Accretion Method, \texttt{CAM}) which connects the observed photospheric composition of early-type stars to the composition of, and accretion from, circumstellar material. We discuss applications to stars hosting disks and planets.
\begin{enumerate}
\item{Equations~\eqref{eq:fraction}, \eqref{eq:abundance}, and \eqref{eq:abundance2} directly relate the properties of the photosphere to both the accretion rate and the observed stellar composition.}
\item{Where the stellar parameters and the accretion rate are observationally constrained, we find very good agreement between the observed photospheric composition and our theoretical predictions. }
\item{We predict that the photospheres of HD~163296, HD~100546, and HD~141569~A are dominated by freshly accreted material. For the latter two, their sub-solar (Fe/H)$_{\rm obs}$ and the presence of dust-poor radial gaps in both disks confirms this. For HD~163296, the observed (Fe/H)$_{\rm obs}$ is slightly super-solar, which is consistent with its accretion from a somewhat gas-poor disk.}
\item{The debris disk host HD~21997 and the hot Jupiter host HD~195689 (KELT-9) are predicted to have photospheres only marginally affected by accretion from these reservoirs, however the peculiar composition of debris disk gas and planetary material may make the contamination detectable for individual elements.}
\end{enumerate}

\label{sec:conclusion}

\section*{Acknowledgements}

The authors are grateful to Quentin Kral, Matteo Cantiello and Simon J. Murphy for helpful comments on this manuscript.
ASJ thanks the UK Marshall commission for financial support. MK gratefully acknowledges funding from the European Union's Horizon 2020 research and innovation programme under the Marie Sklodowska-Curie grant agreement No 753799.

\bibliographystyle{mnras}
\bibliography{refs}

\appendix

\section{Material Flux}
\label{appen:derivative}

Because material is accreting the total derivative left-hand side of equation~\eqref{eq:diff0} is written as
\begin{equation}
\frac{\matd f}{\matd t} = \frac{\partial f}{\partial t} + \frac{\partial z}{\partial t}\frac{\partial f}{\partial z},
\end{equation}
where the partial derivative of $z$ is taken a constant $\Sigma$.
This may equivalently be written as
\begin{equation}
\frac{\matd f}{\matd t} = \frac{\partial f}{\partial t} + \frac{\partial \Sigma}{\partial t}\frac{\partial f}{\partial \Sigma},
\label{eq:total1}
\end{equation}
where the partial derivative of $\Sigma$ is taken at constant $z$.
Defining
\begin{equation}
	\dot{\Sigma} \equiv \frac{\partial \Sigma}{\partial t}
\end{equation}
and inserting equation~\eqref{eq:total1} into equation~\eqref{eq:diff0} we find
\begin{equation}
	\frac{\partial f}{\partial t} + \dot{\Sigma}\frac{\partial f}{\partial \Sigma} = -\frac{1}{\rho}\frac{\partial F}{\partial z}.
	\label{eq:diff1}
\end{equation}
Now using the definition of the column density we see that
\begin{equation}
	\frac{\partial \Sigma}{\partial z} = \rho
\end{equation}
so
\begin{equation}
	\frac{\partial f}{\partial t} + \dot{\Sigma}\frac{\partial f}{\partial \Sigma} = -\frac{\partial F}{\partial \Sigma}.
	\label{eq:diff2}
\end{equation}
Using equation~\eqref{eq:flux0} to eliminate the $\partial f/\partial \Sigma$ we find
\begin{equation}
	\frac{\partial f}{\partial t} - \dot{\Sigma}\frac{F}{\rho^2 D} = -\frac{\partial F}{\partial \Sigma},
	\label{eq:diff3p}
\end{equation}
which is the desired result.

\section{Derivation of $\Sigma_0$}
\label{appen:asymptote}

We are searching for a solution to equation~\eqref{eq:mass}, so
\begin{align}
	\Sigma_{\mathrm{acc}} &= \int_0^{\Sigma_{0}} f(\Sigma) d\Sigma.
\end{align}
Inserting equation~\eqref{eq:f2} we find
\begin{align}
\Sigma_{\mathrm{acc}} &= \int_0^{\Sigma_{0}} 1 - \frac{e^{\dot{\Sigma}\alpha(\Sigma)}-e^{\dot{\Sigma}\alpha(0)}}{e^{\dot{\Sigma}\alpha(\Sigma_0)}-e^{\dot{\Sigma}\alpha(0)}} d\Sigma.
\end{align}
For notational convenience let
\begin{align}
	\gamma(\Sigma) \equiv \alpha(\Sigma)\dot{\Sigma}.
\label{eq:gamma}
\end{align}
Then
\begin{align}
\Sigma_{\mathrm{acc}} &= \int_0^{\Sigma_{0}} 1 - \frac{e^{\gamma(\Sigma)}-e^{\gamma(0)}}{e^{\gamma(\Sigma_0)}-e^{\gamma(0)}} d\Sigma.
\end{align}

Differentiating with respect to $\Sigma_0$ yields
\begin{align}
\frac{d\Sigma_{\mathrm{acc}}}{d\Sigma_0} &= \frac{d}{d\Sigma_0}\int_0^{\Sigma_{0}} 1 - \frac{e^{\gamma(\Sigma)}-e^{\gamma(0)}}{e^{\gamma(\Sigma_0)}-e^{\gamma(0)}} d\Sigma\\
 &= \int_0^{\Sigma_{0}} \frac{d}{d\Sigma_0}\left[1 - \frac{e^{\gamma(\Sigma)}-e^{\gamma(0)}}{e^{\gamma(\Sigma_0)}-e^{\gamma(0)}}\right] d\Sigma,
\end{align}
where we have made use of the fact that the integrand vanishes at $\Sigma_0$.
Evaluating the derivative inside the integral we find
\begin{align}
\frac{d\Sigma_{\mathrm{acc}}}{d\Sigma_0} &= -\int_0^{\Sigma_{0}} \frac{d}{d\Sigma_0}\left[\frac{e^{\gamma(\Sigma)}-e^{\gamma(0)}}{e^{\gamma(\Sigma_0)}-e^{\gamma(0)}}\right] d\Sigma\\
 &= \int_0^{\Sigma_{0}} \frac{d \gamma(\Sigma_0)}{d\Sigma_0}\frac{e^{\gamma(\Sigma_0)}}{e^{\gamma(\Sigma_0)}-e^{\gamma(0)}}\left[\frac{e^{\gamma(\Sigma)}-e^{\gamma(0)}}{e^{\gamma(\Sigma_0)}-e^{\gamma(0)}}\right] d\Sigma\\
 &= \frac{d \gamma(\Sigma_0)}{d\Sigma_0}\frac{e^{\gamma(\Sigma_0)}}{e^{\gamma(\Sigma_0)}-e^{\gamma(0)}}\int_0^{\Sigma_{0}} \frac{e^{\gamma(\Sigma)}-e^{\gamma(0)}}{e^{\gamma(\Sigma_0)}-e^{\gamma(0)}} d\Sigma\\
 &= \frac{d \gamma(\Sigma_0)}{d\Sigma_0}\frac{e^{\gamma(\Sigma_0)}}{e^{\gamma(\Sigma_0)}-e^{\gamma(0)}}\left[\Sigma_0 - \Sigma_{\mathrm{acc}}\right]\\
 &= \frac{d \gamma(\Sigma_0)}{d\Sigma_0}\frac{1}{1-e^{\gamma(0)-\gamma(\Sigma_0)}}\left[\Sigma_0 - \Sigma_{\mathrm{acc}}\right].
\end{align}

Because $\rho$ and $D$ are positive quantities, $\alpha$ is monotonic and hence either asymptotes to a finite value or else diverges.
Because $\dot{\Sigma}$ is also positive, $\gamma$ does the same.
In the event that $\gamma$ diverges we may set $e^{\gamma(0)-\gamma(\Sigma_0)}$ to zero.
Likewise we will argue later that under all cases of interest where $\gamma$ asymptotes, it does so to a value much greater than $\gamma(0)$, and so we may make the same approximation.
Thus we write in both cases
\begin{align}
\frac{d\Sigma_{\mathrm{acc}}}{d\Sigma_0} &\approx \frac{d \gamma(\Sigma_0)}{d\Sigma_0}\left[\Sigma_0 - \Sigma_{\mathrm{acc}}\right],
\end{align}
which may also be written as
\begin{align}
\frac{d\Sigma_{\mathrm{acc}}}{d\Sigma_0} &\approx \frac{d \gamma}{d\Sigma}\bigg|_{\Sigma_0}\left[\Sigma_0 - \Sigma_{\mathrm{acc}}\right].
\label{eq:acc0}
\end{align}
From this we see that it is not the asymptotic behaviour of $\gamma$ that we care about, but rather that of
\begin{align}
\frac{d\gamma}{d\Sigma} = \frac{\dot{\Sigma}}{\rho^2 D}.
\end{align}
Noting that $\rho$ is monotonic in $\Sigma$, and taking $D$ to be monotonic in $\rho$, we find that this either goes to zero, diverges, or approaches a non-zero finite asymptote.

In the event that $d\gamma/d\Sigma\rightarrow \infty$ as $\Sigma \rightarrow \infty$, the solution to equation~\eqref{eq:acc0} asymptotically approach a scenario in which
\begin{equation}
	\Sigma_{0} - \Sigma_{\mathrm{acc}} \approx \frac{d\Sigma}{d\gamma}\bigg|_{\Sigma_0} \rightarrow 0.
\end{equation}
This is because the prefactor of $\Sigma_{\mathrm{acc}}$ on the right-hand side of equation~\eqref{eq:acc0} is negative and so causes the solution to damp towards this point.
Thus in this situation
\begin{equation}
	\Sigma_{\mathrm{acc}} \approx \Sigma_0
	\label{eq:increasingAppen}
\end{equation}
and
\begin{equation}
	f(\Sigma < \Sigma_0) \approx 1.
	\label{eq:fIncreasing}
\end{equation}
Indeed this solution holds approximately even when $d\gamma/d\Sigma$ tends to a non-zero finite asymptote, because in that case the solution tends to
\begin{equation}
	\Sigma_{\mathrm{acc}} = \Sigma_0 - \frac{d\Sigma}{d\gamma}\bigg|_{\Sigma_0} \approx \Sigma_0,
\end{equation}
again with $f(\Sigma < \Sigma_0) \approx 1$.

In the opposite limit, where $d\gamma/d\Sigma\rightarrow 0$ as $\Sigma \rightarrow \infty$, $\Sigma_0$ increases much more rapidly than $\Sigma_\mathrm{acc}$.
Equation~\eqref{eq:acc0} may then be approximated as
\begin{align}
\frac{d\Sigma_{\mathrm{acc}}}{d\Sigma_0} &\approx \frac{d \gamma}{d\Sigma}\bigg|_{\Sigma_0} \Sigma_0.
\label{eq:acc1}
\end{align}
This yields a lower bound on $\Sigma_{\mathrm{acc}}$ of
\begin{equation}
	\Sigma_{\mathrm{acc}} > \frac{1}{2} \frac{d \gamma}{d\Sigma}\bigg|_{\Sigma_0} \Sigma_0^2,
\end{equation}
where we have simply used the derivative evaluated at $\Sigma_0$, which is the minimum value it takes over the interval from $0$ to $\Sigma_0$.

To derive more than a lower bound we need to know more about the structure of $\gamma$.
Under the assumption of hydrostatic equilibrium,
\begin{align}
	\frac{dP}{dr} = -\rho g.
\end{align}
Inserting the ideal gas law we find
\begin{align}
	\frac{d}{dr}\left[\rho T \frac{k_B}{\mu}\right] = -\rho g,
\end{align}
where $\mu$ is the mean molecular weight.
In the outer regions of the star $T$, $g$ and $\mu$ are approximately constant\footnote{See \citet{1969AcA....19....1P} for a more detailed discussion of these approximations.} so
\begin{align}
	\frac{d \ln \rho}{dr} = -\frac{\mu g}{k_B T}.
\end{align}
As such we let
\begin{align}
	h \equiv \frac{k_B T}{\mu g}
	\label{eq:scale}
\end{align}
be the scale height and write
\begin{equation}
	\rho = \rho_0 \exp(z/h),
\end{equation}
which yields
\begin{align}
	\Sigma = \int_{-\infty}^z \rho dz' = h \rho = h \rho_0 e^{-z/h}.
	\label{eq:sigma}
\end{align}
Combining this with equations~\eqref{eq:alpha} and~\eqref{eq:gamma} we find
\begin{equation}
	\frac{d\gamma}{d\Sigma} = \frac{\dot{\Sigma}}{\rho^2 D} = \frac{h^2}{D} \frac{\dot{\Sigma}}{\Sigma^2}.
\end{equation}
As a result equation~\eqref{eq:acc1} may be written as
\begin{equation}
	\frac{d\Sigma_{\mathrm{acc}}}{d\Sigma_0} = \frac{h^2}{D}\frac{\dot{\Sigma}}{\Sigma_0}.
	\label{eq:ddss}
\end{equation}
In all of the cases discussed in Section~\ref{sec:mixingprocesses}, $D$ is a power-law in $\Sigma$ of the form
\begin{equation}
	D = D_0 \Sigma^{\beta}.
	\label{eq:dpower}
\end{equation}
Note that the asymptotic behaviour of $d\gamma/d\Sigma$ changes at $\beta=-2$, and below this point the limit in which it diverges is applicable.
Thus taking $\beta > -2$ and inserting this into equation~\eqref{eq:ddss} we find
\begin{equation}
	\frac{d\Sigma_{\mathrm{acc}}}{d\Sigma_0} = \frac{h^2}{D_0} \dot{\Sigma}\Sigma^{-1-\beta}.
\end{equation}
Integrating and using $\Sigma_\mathrm{acc}(\Sigma_0=0) = 0$ yields
\begin{align}
	\Sigma_{\mathrm{acc}} &= \frac{h^2 \dot{\Sigma}\Sigma_0^{-\beta}}{(-\beta) D_0}\\
	&=\frac{h^2 \dot{\Sigma}}{|\beta| D(\Sigma_0)},
	\label{eq:decreasingAppen}
\end{align}
so long as $\beta < 0$.

The resulting profile for $f$ depends on depth.
Assuming that $\gamma - \gamma(\Sigma_0) \ll 1$ and neglecting $e^{\gamma(0)}$ as discussed, we find that
\begin{equation}
	f(\Sigma) \approx \gamma(\Sigma_0) - \gamma.
	\label{eq:fDecreasingG}
\end{equation}
At shallower depths $f \approx 1$.
Thus we may write
\begin{equation}
	f(\Sigma) \approx \min\left(1, \gamma(\Sigma_0) - \gamma\right).
\end{equation}

In the event that $D$ is a power-law we can evaluate $\gamma$ to find
\begin{equation}
	f(\Sigma) \approx \min\left(1, \frac{h^2\dot{\Sigma} }{\Sigma D(\Sigma)} - \frac{h^2\dot{\Sigma} }{\Sigma_0 D(\Sigma_0)}\right),
\end{equation}
where we have let $\Sigma_0\rightarrow \infty$ and dropped multiplicative factors depending only on $\beta$.
Note that for $-1 < \beta < 0$, $\gamma(\Sigma_0)$ and hence $(\Sigma_0 D(\Sigma_0))^{-1}$ vanishes asymptotically and so at long times we may neglect this term and find the time-independent result that
\begin{equation}
	f(\Sigma) \approx \min\left(1, \frac{h^2\dot{\Sigma} }{\Sigma D(\Sigma)}\right).
\end{equation}
For $-2 < \beta \leq -1$, $\gamma(\Sigma_0)$ grows without bound even though its derivative asymptotes to zero.
In this case there is still a time-dependence to the profile even at long times.
When $\beta = -1$, a case which is both physically relevant and mathematically convenient,
\begin{equation}
	f(\Sigma, t) \approx 1 - \left(\frac{\Sigma}{\Sigma_0}\right)^{h^2 \dot{\Sigma}/D_0}
\end{equation}
and
\begin{equation}
	\Sigma_\mathrm{acc} = \frac{h^2 \dot{\Sigma}}{D_0} \Sigma_0,
\end{equation}
hence
\begin{equation}
	\Sigma_0 = \frac{D_0}{h^2} t.
\end{equation}
Letting
\begin{equation}
	t_{\mathrm{mix}}(\Sigma) \equiv \frac{h^2}{D_0} \Sigma
\end{equation}
we find
\begin{equation}
	f(\Sigma, t) \approx 1 - \left(\frac{t_\mathrm{mix}(\Sigma)}{t}\right)^{h^2 \dot{\Sigma}/D_0},
\end{equation}
which in a series expansion yields
\begin{equation}
	f(\Sigma, t) \approx \frac{h^2 \dot{\Sigma}}{D_0} \ln \frac{t}{t_\mathrm{mix}} = \frac{h^2 \dot{\Sigma}}{D(\Sigma) \Sigma} \ln \frac{t}{t_\mathrm{mix}}.
	\label{eq:fBetaMinus1}
\end{equation}
This form is valid so long as it yields $f < 1$, at which point the appropriate approximation is $f = 1$.
For further decreases in $\beta$ the time-dependence becomes stronger, but in all cases we have considered thus far we find a number well-approximated by
\begin{equation}
	f(\Sigma) \approx \min\left(1, \frac{h^2\dot{\Sigma} }{\Sigma D(\Sigma)}\right),
\end{equation}
which is at most wrong by logarithmic factors in time.

The only case not yet considered is $\beta > 0$, in which $d\gamma/d\Sigma$ goes to zero rapidly as $\Sigma \rightarrow \infty$.
In this limit material is brought so rapidly into the star that it is no longer valid to treat the centre as being infinitely far away from the surface.
That is, the interior boundary condition matters in this limit.
This does not occur in any of the physical cases considered, and so we need not consider it further.

In all cases note that we have found $\Sigma_{0} > \Sigma_{\mathrm{acc}}$, which must be the case because $f \leq 1$ everywhere.
As a result, however, once we have waited a long time relative to the time required to accrete material corresponding to some depth $\Sigma$, we can be assured that $\Sigma_0$ exceeds this depth.
This justifies our assumption that $\Sigma_0$ is large relative to photospheric depths, as even $10^{-15}M_\odot\,\mathrm{yr}^{-1}$ accreted for $30\mathrm{yr}$ suffices to produce $\Sigma_{\mathrm{acc}}$ well in excess of the typical photospheric depths of $0.1\mathrm{g\,cm^{-2}}$. 

Finally, earlier in this section we claimed that the asymptotic value of $\gamma(\Sigma_0)-\gamma(0)$ is, in the cases of interest, generically large when finite.
We now justify this claim.
First suppose that $D$ is a power-law in $\Sigma$ as in equation~\eqref{eq:dpower}.
When $\beta > -1$, $\gamma(0) \rightarrow -\infty$ while $\gamma(\Sigma_0)\rightarrow$ goes to a constant as $\Sigma_0\rightarrow \infty$, so $\gamma(\Sigma_0)-\gamma(0)\rightarrow \infty$.
When $\beta < -1$, $\gamma(0)$ is finite but $\gamma(\Sigma_0) \propto \Sigma_0^{-1-\beta} \rightarrow \infty$, so once more $\gamma(\Sigma_0)-\gamma(0)\rightarrow \infty$.
This covers all mixing mechanisms we have considered in this paper.
Note that even if we regularise the integrals which have $\beta > -1$~\footnote{This would be accomplished by cutting off the integration at some shallow point, which could be physically motivated if the accretion has a significant radial component or otherwise mixes immediately to some non-trivial depth.}, the regularisation point must be shallower than the photosphere~\footnote{Otherwise there is no point in the calculation because the observed $f$ will be unity.}, where the densities are extremely small, and so $\gamma$ still ought to vary substantially.
As a result we expect that $\gamma$ may be taken to have a large dynamic range between the point at which material enters the system and the deepest regions.

Suppose though that we are interested in a case where this argument fails, as there are pathological cases in which it does, such as the limit of vanishing $\dot{\Sigma}$ with a regularised integral.
We would then have $\gamma(0)$ close to the asymptotic value of $\gamma(\Sigma_0)$.
Then expanding equation~\eqref{eq:f2} we find
\begin{align}
	f \approx \frac{\gamma(\Sigma_0) - \gamma(\Sigma)}{\gamma(\Sigma_0) - \gamma(0)}.
\end{align}
which is enhanced relative to equation~\eqref{eq:fincrease} by a factor of $(\gamma(\Sigma_0) - \gamma(0))^{-1}$.
This means that we ought to infer $\Sigma_0$ be smaller by the same factor for a fixed accreted mass, and also means that the profile of $f$ near the surface becomes independent of $\dot{\Sigma}$, which only enters multiplicatively in $\gamma$.
This is physically what happens when the accretion rate is low and the dynamic diffusivity increases rapidly inwards, as then material is then whisked away on a timescale which is much faster than the rate at which it accumulates, so the diffusivity at the point where it arrives is the limiting factor.
This scenario is somewhat artificial, in that all mechanisms we have examined lack the regularising cutoff and hence do not exhibit such behaviour, but it is useful to keep in mind in the event that such a cutoff becomes physically interesting.
This usually occurs at higher accretion rates, where the radial velocity is significant, but there could be cases we have not considered in which it becomes relevant.

\section{Quasi-Steady State Assumption}
\label{appen:steady}

To validate the assumption that mixing processes may be treated as being in a quasi-steady state (i.e. an instantaneous equilibrium) we differentiate equations~\eqref{eq:fIncreasing} and~\eqref{eq:fDecreasingG} in time.
The former is trivial, and yields zero everywhere except near $\Sigma_0$, where the steady-state approximation must break regardless because this is the location of the diffusion front.
The latter is trivial when $f \approx 1$, and yields
\begin{equation}
	\frac{d f}{d t} = \frac{d \gamma(\Sigma_0)}{d t},
\end{equation}
where we have used the fact that only $\Sigma_0$ depends on time to evaluate this derivative.
Now
\begin{align}
\frac{d \gamma(\Sigma_0)}{d t} &= \left.\frac{d \gamma}{d\Sigma}\right\rvert_{\Sigma_0}\frac{d \Sigma_0}{d t}\\
&= \left.\frac{d \gamma}{d\Sigma}\right\rvert_{\Sigma_0}\frac{d\Sigma_0}{d\Sigma_{\mathrm{acc}}}\frac{d\Sigma_\mathrm{acc}}{d t}\\
&= \dot{\Sigma}\left.\frac{d \gamma}{d\Sigma}\right\rvert_{\Sigma_0}\frac{d\Sigma_0}{d\Sigma_{\mathrm{acc}}}.
\end{align}
Inserting equation~\eqref{eq:acc0} we find
\begin{align}
\frac{d \gamma(\Sigma_0)}{d t} \approx \frac{\dot{\Sigma}}{\Sigma_0-\Sigma_{\mathrm{acc}}}.
\end{align}
We are working in the limit where $d\gamma/d\Sigma\rightarrow 0$, so $\Sigma_0 \gg \Sigma_{\mathrm{acc}}$, and hence
\begin{align}
\frac{df}{dt} = \frac{d \gamma(\Sigma_0)}{d t} \approx \frac{\dot{\Sigma}}{\Sigma_0}.
\end{align}
In order to neglect this it must be smaller than the term $dF/d\Sigma$ on the right-hand side of equation~\eqref{eq:diff3}.
In other words, we need the dimensionless ratio
\begin{align}
	\lambda \equiv \frac{d f/d t}{dF/d\Sigma}
\end{align}
to be small.
Evaluating this yields
\begin{align}
	\lambda &=  \left(\frac{\dot{\Sigma}}{\Sigma_0}\right)\left(\frac{dF}{d\Sigma}\right)^{-1}\\
&= \left(\frac{\dot{\Sigma}}{\Sigma_0}\right)\left(\frac{d}{d\Sigma}\left[-\rho^2 D \frac{\partial f}{\partial \Sigma}\right]\right)^{-1}\\
&\approx \left(\frac{\dot{\Sigma}}{\Sigma_0}\right)\left(\frac{d}{d\Sigma}\left[\rho^2 D\frac{d\gamma}{d\Sigma}e^{\gamma-\gamma(\Sigma_0)}\right]\right)^{-1}\\
&= \left(\frac{\dot{\Sigma}}{\Sigma_0}\right)\left(\frac{d}{d\Sigma}\left[\dot{\Sigma}e^{\gamma-\gamma(\Sigma_0)}\right]\right)^{-1}\\
&= \left(\frac{\dot{\Sigma}}{\Sigma_0}\right)\left(\dot{\Sigma}\frac{d\gamma }{d\Sigma}e^{\gamma-\gamma(\Sigma_0)}\right)^{-1}\\
&= \left(\frac{\dot{\Sigma}}{\Sigma_0}\right)\left(\frac{\dot{\Sigma}^2}{\rho^2 D}e^{\gamma-\gamma(\Sigma_0)}\right)^{-1}\\
&= \left(\frac{\rho^2 D}{\dot{\Sigma}\Sigma_0}\right)\left(e^{\gamma-\gamma(\Sigma_0)}\right)^{-1}.
\end{align}
Noting that $e^{\gamma-\gamma(\Sigma_0)} \approx 1$ in this regime, we find
\begin{align}
	\lambda &\approx \frac{\rho^2 D}{\dot{\Sigma}\Sigma_0}\\
&\approx \frac{\Sigma^2 D}{h^2 \dot{\Sigma} \Sigma_0},
\end{align}
where we have made use of equation~\eqref{eq:sigma} in the last line.
Combining this with equation~\eqref{eq:decreasingAppen} we find
\begin{align}
	\lambda \approx |\beta|^{-1}\frac{\Sigma^2}{\Sigma_{\mathrm{acc}}\Sigma_0} \approx \frac{\Sigma^2}{\Sigma_{\mathrm{acc}}\Sigma_0},
\end{align}
where we have dropped $\beta$ because it is of order unity.
Thus $\lambda$ is small in the regime where
\begin{equation}
	\Sigma \ll \sqrt{\Sigma_0 \Sigma_{\mathrm{acc}}}.
\end{equation}
This is as expected: the approximation is a good one certainly up to $\Sigma_{\mathrm{acc}}$ and for much of the region beyond that, but breaks down as we approach $\Sigma_0$, which is the location of the diffusion front.

In summary, the region near the surface of the star is the one of most interest, and in all cases considered above we have found that for $\Sigma \ll \Sigma_{\mathrm{acc}}$ the quasi-steady state approximation holds as desired.

\section{Opacities}
\label{appen:opac}

To determine the opacity of bulk stellar material we used the OPAL~\citep{1996ApJ...464..943I} and Ferguson~\citep{0004-637X-623-1-585} opacity tables for solar composition and metallicity.
This is not quite right when the composition is different from solar, but because this is only used to determine the depth of the photosphere it does not contribute a significant error in the calculations of chemical mixing.
The Ferguson tables are more accurate at low temperatures while the OPAL ones extend to much higher temperatures, so we choose the Ferguson opacities where available and the OPAL ones otherwise.

\section{Convective Diffusivity}
\label{appen:conv}

The convective diffusivity is given by equation~\eqref{eq:conv} as
\begin{align}
	D \approx h \left(\frac{F_\star}{\rho}\right)^{1/3}.
\end{align}
This may also be written as
\begin{align}
	D \approx h c_\mathrm{s} \left(\frac{F_\star}{\rho c_\mathrm{s}^3}\right)^{1/3}.
\end{align}
Noting that $P \approx \rho c_\mathrm{s}^2$ we find that
\begin{align}
	D \approx h c_\mathrm{s} \left(\frac{F_\star}{P c_\mathrm{s}}\right)^{1/3}.
\end{align}
We are interested in the diffusivity at the photosphere, which is given by equation~\eqref{eq:depth1} as
\begin{equation}
	P \approx g \kappa^{-1},
\end{equation}
so
\begin{align}
	D \approx h c_\mathrm{s} \left(\frac{F_\star \kappa}{g c_\mathrm{s}}\right)^{1/3}.
\end{align}
The sound speed is proportional to $T^{1/2}$, the scale height to $T/g$, and the flux to $T^4$, so neglecting the effects of molecular weight variations
\begin{align}
	D \approx h_\odot c_{\mathrm{s},\odot} \left(\frac{F_\odot \kappa_\odot}{g_\odot c_{\mathrm{s},\odot}}\right)^{1/3} \left(\frac{T}{T_\odot}\right)^{8/3}\left(\frac{g}{g_\odot}\right)^{-4/3} \left(\frac{\kappa}{\kappa_\odot}\right)^{1/3}.
\end{align}
Making use of $L \propto M^{3.5}$ and $R \propto M$~\citep{1992isa..book.....B} we find that $T^4 \propto M^{3/2}$ and hence $T^{8/3} \propto M$.
Similarly we find that $g \propto M^{-1}$, and hence
\begin{align}
	D \approx h_\odot c_{\mathrm{s},\odot} \left(\frac{F_\odot \kappa_\odot}{g_\odot c_{\mathrm{s},\odot}}\right)^{1/3} \left(\frac{M}{M_\odot}\right)^{7/3} \left(\frac{\kappa}{\kappa_\odot}\right)^{1/3}.
\end{align}
Using $\kappa_\odot \approx 10\mathrm{cm^2/g}$ we find
\begin{align}
	D \approx 7\times 10^{13}\mathrm{cm^2 g^{-1}}\left(\frac{M}{M_\odot}\right)^{7/3} \left(\frac{\kappa}{10\mathrm{cm^2 g^{-1}}}\right)^{1/3}.
\end{align}

\bsp
\label{lastpage}
\end{document}